# Get the Most out of Your Sample: Optimal Unbiased Estimators using Partial Information

Edith Cohen
AT&T Labs–Research
180 Park Avenue
Florham Park, NJ 07932, USA
edith@research.att.com

Haim Kaplan
School of Computer Science
Tel Aviv University
Tel Aviv, Israel
haimk@cs.tau.ac.il

## ABSTRACT

[1] Random sampling is an essential tool in the processing and transmission of data. It is used to summarize data too large to store or manipulate and meet resource constraints on bandwidth or battery power. Estimators that are applied to the sample facilitate fast approximate processing of queries posed over the original data and the value of the sample hinges on the quality of these estimators.

Our work targets data sets such as request and traffic logs and sensor measurements, where data is repeatedly collected over multiple *instances*: time periods, locations, or snapshots. We are interested in queries that span multiple instances, such as distinct counts and distance measures over selected records. These queries are used for applications ranging from planning to anomaly and change detection.

Unbiased low-variance estimators are particularly effective as the relative error decreases with the number of selected record keys. The Horvitz-Thompson estimator, known to minimize variance for sampling with "all or nothing" outcomes (which reveals exacts value or no information on estimated quantity), is not optimal for multi-instance operations for which an outcome may provide partial information.

We present a general principled methodology for the derivation of (Pareto) optimal unbiased estimators over sampled instances and aim to understand its potential. We demonstrate significant improvement in estimate accuracy of fundamental queries for common sampling schemes.

## 1. INTRODUCTION

Random sampling had become an essential tool in the handling of data. It is used to accommodate resource constraints on storage, bandwidth, energy, and processing power. Massive data sets can be too large to be stored long term or transmitted, sensor nodes collecting measurements are energy limited, and even when the full data is available, computation of exact aggregates may be slow and costly.

The sample constitutes a summary of the original data sets that is small enough to store, transmit, and manipulate in a single location and yet supports computation of approximate queries over the original data. It is flexible in that many types of queries are supported and that queries need not be known a priori [31, 39, 5, 4, 9, 25, 26, 2, 21, 2, 27, 13, 22, 10, 14].

Commonly, data has the form of multiple *instances* which are *dispersed* in time or location. Each instance corresponds to an assignment of values to a set of identifiers (keys). The universe of key values is shared between instances but the values change. This data can be modeled as a numeric matrix of instances $\times$ keys. Instances can be snapshots of a database that is modified over time, measurements from sensors or of parameters taken in different time periods, or number of requests for resources processed at multiple servers. Clearly, any scalable summarization algorithm of dispersed data must decouple the processing of different instances: the processing of one instance must not depend on values in other instances.

[1]This is a full version of [15].

An important class of query primitives are functions with arguments that span values assumed by a key in multiple instances, such as quantiles (maximum, minimum, median) or the *range* (difference between maximum and minimum). *Sum aggregates* of these primitives over selected subsets of keys [32, 8, 17] include distinct element count (size of union), max-dominance and min-dominance norms [19, 20] and the Manhattan ($L_1$) distance and are used for change or anomaly detection, similarity-based clustering, monitoring, and planning. See example in Figure 5 (A).

Popular sampling scheme of a single instance are Poisson – where keys are sampled independently, bottom-$k$ (order) [36, 12, 22, 13, 14] – where keys are assigned random rank values and the $k$ smallest ranked keys are selected (as in weighted sampling without replacement and priority sampling), and VarOpt [10, 6] .

The Horvitz Thompson (HT) estimator [29], based on inverse-probability weights, is a classic method for estimating subset-sums of values of keys: The estimate on the value of a key is 0 if it is not included in the sample and the ratio of its true value and the inclusion probability otherwise. The estimate on the sum of values of a subset of keys is the sum of estimates over sampled keys that are members of the subset. This estimator is unbiased and has minimum variance amongst unbiased nonnegative estimators. A variant of HT is used for bottom-$k$ sampling [22, 38, 17].

Previous estimators we are aware of for multi-instance functions are based on an adaptation of HT: a positive estimate is provided only on samples that revealed sufficient information to compute the exact value of the estimated quantity. We observe that such estimators may not be optimal for multi-instance functions, where outcomes can provide partial information on the estimated value. We aim to understand the form and potential performance gain of better estimators.

**Contribution:** We characterize the joint sample distributions attainable for dispersed instances, that is, when processing of each instance may not depend on values of another.

Our main contribution is a principled methodology for deriving optimal estimators for multi-instance functions, taking the sampling scheme as a given. The sample of each instance can be Poisson, VarOpt, or bottom-$k$. Sampling can be weighted (inclusion probability in the sample depends on the value) or weight-oblivious. The joint distribution (samples of different instances) can be *independent* or *coordinated*. Coordination, achieved using random hash function, means that similar instances get similar samples[3, 37, 34, 36, 5, 4, 9, 25, 26, 2, 13, 27, 14, 17] and can boost estimation quality of multi-instance functions [17, 18].

We provide example derivations of optimal estimators for basic aggregations over common sampling distributions and demonstrate significant gain, in terms of lower variance, over state-of-the-art estimators. Optimality is in a Pareto sense with respect to variance:

any other nonnegative estimator with lower variance on some data must have higher variance on some other data.

A key component in attaining optimality is the use of partial information, which we motivate by the following simple scenario. Consider estimating the maximum of two values, $v_1$ and $v_2$, sampled independently with respective probabilities $p_1$ and $p_2$. We can be certain about the value $\max(v_1, v_2)$ only when both values are sampled, which happens with probability $p_1 p_2$. The inverse-probability weight is $\max(v_1, v_2)/(p_1 p_2)$ when both values are sampled and 0 otherwise and is an unbiased estimate. We now observe that when exactly one of the values is sampled, we know that the maximum is at least that value, that is, we have meaningful partial information in the form of a positive lower bound on the maximum. We will show how to exploit that and obtain a nonnegative and unbiased estimator with lower variance than the inverse-probability weight.

We distinguish between independent weighted sampling schemes, according to the "reproducibility" of the randomization used: with *Known (unknown) seeds* the random hash functions used in sampling each instance are (are not) available to the estimator. We show that knowledge of seeds substantially increases estimation power: we provide nonnegative unbiased estimators for the maximum when seeds are known and show that when seeds are unknown, there is no such estimator even when there are only two values and the domain is Boolean (in which case the maximum is OR of two bits). Our negative result for unknown seeds agrees with prior work that (implicitly) assume "unknown seeds," such as [7], who showed that most of the data needs to be sampled in order to obtain with constant probability small error estimate of distinct element count (which is a sum aggregate of OR). "Known seeds" sampling, however, can be easily incorporated when streaming or otherwise processing the full data set. We demonstrate its benefit when independent weighted samples of instances might be used post hoc for estimates of multi-instance queries. While reproducible randomization was extensively used as a means to coordinate samples, we believe that its potential to enhance the usefulness of independent weighted samples was not previously properly understood.

**Overview:** Section 2 characterizes all sample distributions that are consistent with the constraints on summarization of dispersed values. In Section 3 we propose methods to obtain optimal estimators which we apply in Sections 4-5 in example derivations. In Section 4 we consider weight-oblivious Poisson sampling of keys and independent sampling of instances and derive two Pareto optimal estimators for the maximum, one catering for data where values of a key are similar across instances and one where variation is large. Weighted sampling (with known seeds) is studied in Section 5 and we derive optimal estimators for the maximum and Boolean OR over two instances. Section 6 contains negative results for independently sampled instances with unknown seeds: We show that there are no unbiased nonnegative estimators for maximum and for absolute difference, even when data is binary.

In terms of an instances $\times$ keys data matrix, Sections 2–6 consider functions over the values $\mathbf{v} = (v_1, \ldots, v_r)$ of a *single* key (i.e., column) in $r$ dispersed instances. To estimate sum aggregates over multiple selected keys, we sum individual estimates for the selected keys. For example, to estimate distinct element count, we apply an OR estimator for each key and sum these estimates. Section 7 overviews the application of single-key estimators to sum aggregates. Applications to distinct count and max dominance are provided in Section 8.

# 2. SAMPLING DISPERSED VALUES

The data is represented by a vector $\mathbf{v} = (v_1, \ldots, v_r) \in \mathbf{V}$ where $\mathbf{V} \subset V_1 \times \cdots \times V_r$ and we are interested in the value of a function $f(\mathbf{v})$. Examples include the value $v_i$ of the $i$th entry, the $\ell$th largest entry $\ell^{\text{th}}(\mathbf{v})$, the maximum $\max(\mathbf{v}) = \max_{i \in [r]} v_i$, the minimum $\min(\mathbf{v}) = \min_{i \in [r]} v_i$, the range $\text{RG}(\mathbf{v}) = \max(\mathbf{v}) - \min(\mathbf{v})$, and exponentiated range $\text{RG}_d(\mathbf{v}) \equiv \text{RG}(\mathbf{v})^d$ for $d > 0$. The domain $\mathbf{V}$ can be the nonnegative quadrant of $\mathbb{R}^r$ or $\{0,1\}^r$.

For a subset $V' \subset \mathbf{V}$ of data vectors we define $\underline{f}(V') = \inf\{f(v) \mid v \in V'\}$ and $\overline{f}(V') = \sup\{f(v) \mid v \in V'\}$, the lowest and highest values of $f$ on $V'$.

We see a random sample $S \subset [r]$ of the entries of $\mathbf{v}$. The sample distribution is subject to the constraint that the inclusion of $i$ in $S$ is independent of the values $v_j$ for $j \neq i$. This is formalized as follows: There is a probability distribution $\mathcal{T}$ over a sample space $\Omega$ of predicates $\boldsymbol{\sigma} = (\sigma_1, \ldots, \sigma_r)$, where $\sigma_i$ has domain $V_i$. The sample $S \equiv S(\boldsymbol{\sigma}, \mathbf{v})$ is a function of the predicate vector $\boldsymbol{\sigma}$ and the data vector $\mathbf{v}$ and includes $i \in [r]$ if and only if $\sigma_i(v_i)$ is true: $i \in S \Leftrightarrow \sigma_i(v_i)$.

Two special cases are:

- *Weight-oblivious* sampling, where inclusion of $i$ in $S$ is independent of $v_i$. The predicates $\sigma_i$ are constants (0 or 1) and entry $i$ is sampled if and only if $\sigma_i = 1$ (which happens with probability $p_i = \mathsf{E}[\sigma_i]$).
- Weighted sampling where inclusion probability of each $i$ is nondecreasing with $v_i$ (in particular, $v_i = 0 \implies i \notin S$).

Weighted sampling is important when the sample is used to estimate functions that increase with the data values. The predicates $\sigma_i$ are increasing functions that can be specified in terms of a transition threshold value $\tau_i$:

$$i \in S \iff \sigma_i(v_i) \iff v_i \geq \tau_i \ .$$

We find it convenient to specify weighted sampling distributions using non-decreasing functions $\tau_i : [0,1]$, $i \in [r]$ and a random *seed* vector $\mathbf{u} \in [0,1]^r$ so that $u_i \in [0,1]$ is uniformly distributed, with the interpretation that

$$i \in S \iff v_i \geq \tau_i(u_i) \ .$$

The inclusion probability of $i$ is $\text{PR}[v_i \geq \tau_i(u_i)] = \sup\{u \in [0,1] \mid v_i \geq \tau_i(u)\}$.

Weighted sampling is *PPS* (Probability Proportional to Size) when $\tau_i = u_i \tau_i^*$, where $\boldsymbol{\tau}^*$ is a fixed vector. With PPS sampling, $i$ is sampled with probability $\min\{1, v_i/\tau_i^*\}$.

**Independent (Poisson)** sampling is when entries are sampled independently, that is, the seeds $u_i$ are independent. In the general model, $\mathcal{T}$ is a product distribution and $\sigma_i$ is independent of all $\sigma_j$ for $j \neq i$.

**Shared-seed (coordinated) sampling** is when the entries of the seed vector are identical: $u_1 = \cdots = u_r \equiv u$ where $u \in [0,1]$ is selected uniformly at random.

## 2.1 Estimators

An estimator $\hat{f}(S)$ of $f(\mathbf{v})$ is a function applied to the outcome $S$ (sampled entries and their values). The estimator depends on the domain $\mathbf{V}$ and distribution $\mathcal{T}$. When sampling is weighted, we distinguish between two models, depending whether the *seeds* (the random predicate vector $\boldsymbol{\sigma}$ in the general model or the seed vector $\mathbf{u}$) are available to the estimator. From the seed we can reveal information on values of entries that are not included in the sample: If $i \notin S$, we know that $v_i < \tau_i(u_i)$ ($v_i \in \sigma_i^{-1}(0)$ in the general model).

With an outcome $S$, we associate a set $V^*(S) \subset \mathbf{V}$ of all data vectors consistent with this outcome. In the discrete case,

$V^*(S) = \{\mathbf{v} \in \mathbf{V} \mid \text{PR}[S \mid \mathbf{v}] > 0\}$. Otherwise, $V^*(S)$ contains all vectors for which the probability density for the outcome is positive. When seeds are not known, $V^*(S)$ contains $\mathbf{v}$ if and only if the probability density of $S(\boldsymbol{\sigma}, \mathbf{v})$ (where $\boldsymbol{\sigma} \in \Omega$) is positive for our outcome $S$. With known seeds, $\boldsymbol{\sigma}$ is available, and hence $\mathbf{v} \in V^*(S)$ if and only if the outcome matches $S(\boldsymbol{\sigma}, \mathbf{v})$.

We seek estimators with some or all of the following properties:

*unbiased*: for all $\mathbf{v}$, $\mathsf{E}[\hat{f} \mid \mathbf{v}] = f(\mathbf{v})$.

*nonnegative*: $\hat{f} \geq 0$.

*bounded variance*: $\forall \mathbf{v}$, $\text{VAR}[\hat{f} \mid \mathbf{v}] < \infty$.

*dominance:* We say that an estimator $\hat{f}^{(1)}$ *dominates* $\hat{f}^{(2)}$ if for all data vectors $\mathbf{v}$, $\text{VAR}[\hat{f}^{(1)} \mid \mathbf{v}] \leq \text{VAR}[\hat{f}^{(2)} \mid \mathbf{v}]$. An estimator $\hat{f}$ is *dominant* (Pareto optimal) if there is no other unbiased nonnegative estimator $\hat{f}'$ that dominates $\hat{f}$.

*monotone*: Nonnegative and non-decreasing with information. If $V^*(S) \subset V^*(S')$, then $\hat{f}(S) \geq \hat{f}(S')$.

Unbiasedness is particularly desirable when estimating sums by summing individual estimates: When unbiased and independent (or non-positively correlated) estimates are combined, the relative error decreases. Nonnegativity is desirable when estimating a nonnegative function $f \geq 0$, ensuring an estimate from the same domain as the estimated quantity. If there is an estimator that dominates all others, it is the only optimal one. If there isn't, we instead aim for Pareto optimality. Monotonicity is an intuitive smoothness requirement.

## 2.2 Horvitz Thompson estimator

Suppose we are interested in estimating a function $f(v) \geq 0$ under "all or nothing" sampling, where either the value is sampled and $v$ is known precisely or it is not sampled and we know nothing about $f(v)$. When the value is sampled, from the value $v$ and the sample distribution we can compute the probability $p$ that the value is sampled.

The HT estimator [29] $\hat{f}^{(HT)}$ of $f(v)$ applies *inverse probability weighting*: $\hat{f} = 0$ if the entry is not sampled and $\hat{f} = f(v)/p$ if it is sampled. This estimator is clearly nonnegative, monotone, and unbiased: $\mathsf{E}[\hat{f}] = (1-p) * 0 + p\frac{f(v)}{p} = f(v)$ . The variance is

$$\text{VAR}[\hat{f}] = f(v)^2 \left( \frac{1}{p} - 1 \right) . \quad (1)$$

The HT estimator is optimal in that $\text{VAR}[\hat{f}]$ is minimized for all $\mathbf{v}$ over all unbiased nonnegative estimators. Intuitively, this is because an unbiased nonnegative estimator can not be positive (with nonzero probability) on outcomes that are consistent with $f(v) = 0$ and variance is minimized when using equal estimate when sampled.

**Multi-entry** $f$. The application of inverse-probability weights on multi-entry functions is more delicate. We can use the set of outcomes for which $S = [r]$, that is all entries are sampled. For these outcomes we know the data $\mathbf{v}$ and from $\mathcal{T}$ we can determine $\text{PR}[S = [r] \mid \mathbf{v}]$. The estimator is $f(\mathbf{v})/\text{PR}[S = [r] \mid \mathbf{v}]$ if $S = [r]$ and 0 otherwise. This estimator is defined when $\text{PR}[S = [r] \mid \mathbf{v}] > 0$. With weighted sampling, however, "0" valued entries are never sampled, so we may have $\text{PR}[S = [r]] = 0$ when $f(\mathbf{v}) > 0$.

A broader definition of inverse-probability estimators [17, 18] is with respect to a subset $\mathcal{S}^*$ of all possible outcomes (over $\Omega$ and $\mathbf{V}$). The outcomes $\mathcal{S}^*$ are those on which the estimator is positive. The estimator is defined for $\mathcal{S}^*$ if there exist two functions $f^*$ and $p^*$ with domain $\mathcal{S}^*$ that satisfy the following:

- for any outcome $S \in \mathcal{S}^*$, for all $\mathbf{v} \in V^*(S)$, $f(\mathbf{v}) = f^*(S)$ and $\text{PR}[\mathcal{S}^* \mid \mathbf{v}] = p^*(S)$.
- for all $\mathbf{v} \in \mathbf{V}$ with $f(v) > 0$, $\text{PR}[\mathcal{S}^* \mid \mathbf{v}] > 0$.

The estimate is $\hat{f}(S) = 0$ if $S \notin \mathcal{S}^*$ and $\hat{f}(S) = f^*(S)/p^*(S)$ otherwise. These functions and hence the estimator are unique for $\mathcal{S}^*$ if they exist. When $\mathcal{S}^*$ is more inclusive, the respective estimator has lower (or same) variance on all data. We use the notation $\hat{f}^{(HT)}$ for the estimator corresponding to the most inclusive $\mathcal{S}^*$. A sufficient condition for optimality of $\hat{f}^{(HT)}$ is that for all outcomes $S \notin S^*$, $\underline{f}(V^*(S)) = 0$.

## 2.3 Necessary conditions for estimation

Inverse-probability estimators are unbiased, nonnegative (when $f$ is) , and monotone. At most two different estimate values (zero and possibly a positive value) are possible for a given data vector and thus, variance is bounded. An inverse-probability estimator, however, exists only if for all data such that $f(\mathbf{v}) > 0$, there is positive probability of recovering $f(\mathbf{v})$ from the outcome. This requirement excludes basic functions such as RG over weighted samples: When the data has at least one positive and one zero entry, there is zero probability of recovering the exact value of $\text{RG}(\mathbf{v})$ from the outcome. A nonnegative, unbiased, and bounded-variance RG estimator, however, was presented in [17, 18].

Aiming for a broader understanding of when an estimator with these properties exists, we derive some necessary conditions. For a set of outcomes, determined by a portion $\Omega' \subset \Omega$ of the sample space and data vector $\mathbf{v}$, we define

$$V^*(\Omega', \mathbf{v}) = \bigcap_{\boldsymbol{\sigma} \in \Omega'} V^*(S(\sigma, \mathbf{v}))$$

the set of all vectors that are consistent with all outcomes determined by $\Omega'$ and $\mathbf{v}$.
For $\mathbf{v}$ and $\epsilon$, we define $\underline{\Delta}(\mathbf{v}, \epsilon) = 1$ if $\forall \boldsymbol{\sigma}, \underline{f}(S(\boldsymbol{\sigma}, v)) > f(\mathbf{v}) - \epsilon$ and

$$\underline{\Delta}(\mathbf{v}, \epsilon) = 1 - \sup \left\{ \text{PR}[\Omega'] \mid \Omega' \subset \Omega,\ \underline{f}(V^*(\Omega', \mathbf{v})) \leq f(\mathbf{v}) - \epsilon \right\} \quad (2)$$

otherwise.

That is, we look for $\Omega'$ of maximum size such that if we consider all vectors $\mathbf{v}' \in V^*(\Omega', \mathbf{v})$ that are consistent with $\mathbf{v}$ on $\Omega'$, the infimum of $f$ over $V^*(\Omega', \mathbf{v})$ is at most $f(\mathbf{v}) - \epsilon$. We define $\underline{\Delta}(\mathbf{v}, \epsilon)$ as the probability $\text{PR}[\Omega \setminus \Omega']$ of not being in that portion.

LEMMA 2.1. *A function $f$ has an estimator that is*

- *unbiased and nonnegative $\Rightarrow$:*

$$\forall \mathbf{v}, \forall \epsilon > 0, \underline{\Delta}(\mathbf{v}, \epsilon) > 0 \quad (3)$$

- *unbiased, nonnegative, and bounded variance $\Rightarrow$:*

$$\forall \mathbf{v},\ \underline{\Delta}(\mathbf{v}, \epsilon) = \Omega(\epsilon^2) . \quad (4)$$

- *unbiased, nonnegative, and bounded $\Rightarrow$:*

$$\forall \mathbf{v},\ \underline{\Delta}(\mathbf{v}, \epsilon) = \Omega(\epsilon) . \quad (5)$$

PROOF. The contribution of $\Omega'$ to the expectation of $\hat{f}$ must not exceed $\underline{f}(V^*(\Omega', \mathbf{v}))$. Because if it does, then $\hat{f}$ must assume negative values for $\mathbf{v}' \in V^*(\Omega', \mathbf{v})$ with minimum $f(\mathbf{v}')$. Considering a maximum $\Omega'$ with $\underline{f}(V^*(\Omega',\ \mathbf{v})) \leq f(\mathbf{v}) - \epsilon$, its contribution to the expectation is at most $f(\mathbf{v}) - \epsilon$ and the contribution of the complement, which has probability $\underline{\Delta}(\mathbf{v}, \epsilon)$, must be at least $\epsilon$.

If $\underline{\Delta}(\mathbf{v},\epsilon)=0$ then this is not possible, so (3) follows. The expectation of the estimator over the complement is at least $\frac{\epsilon}{\underline{\Delta}(\mathbf{v},\epsilon)}$, thus (5) is necessary. The contribution to the variance of that complement is at least

$$\underline{\Delta}(\mathbf{v},\epsilon)\left(\frac{\epsilon}{\underline{\Delta}(\mathbf{v},\epsilon)}-f(\mathbf{v})\right)^2$$

which implies (4) is necessary. $\square$

# 3. PARETO OPTIMAL ESTIMATORS

We formulate sufficient conditions for Pareto optimality, which form the basis of our estimator derivations.

We start by seeking Pareto optimal estimators defined with respect to an order $\prec$ over the set $V$ of all possible data vectors and minimizing variance in an order-respecting way: The variance of the estimator for a data vector $\mathbf{v}$ is minimized conditioned on values it assigned to outcomes consistent with vectors that precede $\mathbf{v}$. This setup naturally yields estimators that are Pareto optimal. Moreover, by selecting an order $\prec$ so that more likely vectors appear earlier, we can tailor the estimator according to properties of the data.

**Order-based optimality $\hat{f}^{(\prec)}$:** The first estimator we present, $\hat{f}^{(\prec)}$, is the solution of a simple set of equations. A solution may not exist, but when it does, it is unique and Pareto optimal.

We map an outcome $S$ to its $\prec$-minimal consistent data vector $\phi(S)\equiv\min_{\prec}V^*(S)$ (we assume it is well defined). We say that $S$ is *determined* by the data vector $\mathbf{v}\equiv\phi(S)$ and that $\mathbf{v}$ is the *determining vector* of $S$. An outcome $S$ *precedes* $\mathbf{v}$ if it is determined by $\mathbf{z}\prec\mathbf{v}$.

For continuous spaces $V$ and $\mathcal{S}$, we extend some assumptions on the mapping $\phi$ from the discrete case: (i) For all $\mathbf{v}$, $\phi^{-1}(\mathbf{v})$ is either empty or has positive probability, (ii) any subset of $\phi^{-1}(\mathbf{v})$ with zero probability for data $\mathbf{v}$ also has zero probability for data $\mathbf{z}\succ\mathbf{v}$, and (iii) any positive-probability set of outcomes consistent with $\mathbf{v}$ and determined by preceding vectors must include a positive probability subset of $\phi^{-1}(\mathbf{z})$ for some $\mathbf{z}\prec\mathbf{v}$.

For each vector $\mathbf{v}$, $\hat{f}^{(\prec)}$ has the same value on all outcomes $\mathcal{S}'\equiv\phi^{-1}(\mathbf{v})$ determined by $\mathbf{v}$. Slightly abusing notation, we define $\hat{f}^{(\prec)}(\mathbf{v})\equiv\hat{f}^{(\prec)}(S)$ for $S\in\mathcal{S}'$ to be that value.

We express $\hat{f}^{(\prec)}(\mathbf{v})$ as a function of $\hat{f}^{(\prec)}$ on the outcomes $\mathcal{S}_0$ that precede $\mathbf{v}$. The dependence on preceding outcomes $\mathcal{S}_0$ is through their contribution $f_0$ to the expectation of the estimate of $f(\mathbf{v})$. The estimate value $\hat{f}^{(\prec)}(\mathbf{v})$ is as follows: If $\textsc{pr}[\mathcal{S}'|\mathbf{v}]=0$ and $f(\mathbf{v})=f_0$, $\hat{f}^{(\prec)}(\mathbf{v})\leftarrow 0$. If $\textsc{pr}[\mathcal{S}'|\mathbf{v}]=0$ and $f(\mathbf{v})\neq f_0$, we declare failure. Else,

$$\hat{f}^{(\prec)}(\mathbf{v})\leftarrow\frac{f(\mathbf{v})-f_0}{\textsc{pr}[\mathcal{S}'|\mathbf{v}]}\,. \tag{6}$$

From the inverse-probability weights principle, this choice of $\hat{f}(S)$ for $S\in\mathcal{S}'$ minimizes the variance $\textsc{var}[\hat{f}|\mathbf{v}]$ for data vector $\mathbf{v}$ conditioned on the values $\hat{f}:\mathcal{S}_0$.

When the order $\prec$ enumerates all data vectors (all data vectors have finite position in the order), we can compute $\hat{f}^{(\prec)}$ algorithmically: Algorithm 1 processes data vectors sequentially in increasing $\prec$ order and computes $\hat{f}(\mathbf{v})$ when $\mathbf{v}$ is processed.

These constraints have no solution when for some $\mathbf{v}$, $f_0<f(\mathbf{v})$ and $\textsc{pr}[\mathcal{S}'|\mathbf{v}]=0$. Moreover, if $f_0>f(\mathbf{v})$, there is no nonnegative solution. When a solution $\hat{f}^{(\prec)}$ is well defined, however, it is unbiased and Pareto optimal

**Algorithm 1** $\hat{f}^{(\prec)}$

**Require:** $\prec$ is an order on $V$
1: $\mathcal{S}_0\leftarrow\emptyset$ ▷ set of processed outcomes
2: $V_0\leftarrow\emptyset$ ▷ set of processed data vectors
3: **while** $V_0\neq V$ **do**
4: $\quad\mathbf{v}\leftarrow\min_{\prec}(V\setminus V_0)$ ▷ A minimum unprocessed vector
5: $\quad f_0\leftarrow\mathsf{E}[\hat{f}^{(\prec)}(S)|\mathcal{S}_0,\mathbf{v}]\textsc{pr}[\mathcal{S}_0|\mathbf{v}]$ ▷ Contribution of preceding outcomes to the estimate of $f(\mathbf{v})$
6: $\quad\mathcal{S}'\leftarrow\{S|\mathbf{v}\in V^*(S)\}\setminus\mathcal{S}_0$ ▷ Unprocessed outcomes consistent with $\mathbf{v}$
7: $\quad$**if** $\textsc{pr}[\mathcal{S}'|\mathbf{v}]=0$ **then**
8: $\qquad$**if** $f(\mathbf{v})\neq f_0$ **then return "failure"** ▷ No unbiased estimator
9: $\qquad$**else**
10: $\qquad\quad\hat{f}\leftarrow 0$
11: $\qquad\quad\forall S\in\mathcal{S}'$, $\hat{f}^{(\prec)}(S)\leftarrow 0$
12: $\quad$**else**
13: $\qquad\hat{f}\leftarrow\frac{f(\mathbf{v})-f_0}{\textsc{pr}[\mathcal{S}'|\mathbf{v}]}$
14: $\quad\forall S\in\mathcal{S}'$, $\hat{f}^{(\prec)}(S)\leftarrow\hat{f}$
15: $\quad V_0\leftarrow V_0\cup\{\mathbf{v}\}$
16: $\quad\mathcal{S}_0\leftarrow\mathcal{S}_0\cup\mathcal{S}'$

LEMMA 3.1. *When $\hat{f}^{(\prec)}$ is well defined, it is unbiased and Pareto optimal.*

PROOF. Pareto optimality: Consider an estimator $\hat{f}^{(\prec)}$ such that for some $\mathbf{v}$, $\hat{f}\neq\hat{f}^{(\prec)}$ on a set of outcomes $\mathcal{D}$ such that $\textsc{pr}[\mathcal{D}|\mathbf{v}]>0$. Let $\mathbf{v}$ be $\prec$-minimal with this property, and let $\mathcal{S}_0$ and $\mathcal{S}'$ be as in our constraints, with respect to $\mathbf{v}$. From definition of $\phi$, the set $\mathcal{D}$ (or a same-probability subset of it) must be contained in $\mathcal{S}'$.

From $\prec$-minimality of $\mathbf{v}$ and our assumptions for continuous spaces, we must have $\mathsf{E}[\hat{f}^{(\prec)}:|\mathcal{S}_0]=\mathsf{E}[\hat{f}|\mathcal{S}_0]$ and hence $\hat{f}^{(\prec)}:\mathcal{S}'\neq\hat{f}:\mathcal{S}'$. The value assigned by $\hat{f}^{(\prec)}$ on the outcomes $\mathcal{S}'$ is the *unique* choice which minimizes the variance of $\mathbf{v}$ subject to $\hat{f}^{(\prec)}:\mathcal{S}_0$, in the sense that any estimator that differs on a positive probability subset of $\mathcal{S}'$ will have strictly higher variance. Hence, $\textsc{var}[\hat{f}|\mathbf{v}]>\textsc{var}[\hat{f}^{(\prec)}|\mathbf{v}]$ and thus $\hat{f}$ can not dominate $\hat{f}^{(\prec)}$.

Unbiasedness follows from the choice of $\hat{f}^{(\prec)}$ on the outcomes $\mathcal{S}'$ in (6) (line 13 of Algorithm 1): $\mathsf{E}[\hat{f}^{(\prec)}]=\mathsf{E}[\hat{f}^{(\prec)}|\mathcal{S}']\textsc{pr}[\mathcal{S}']+\mathsf{E}[\hat{f}^{(\prec)}|\mathcal{S}_0]\textsc{pr}[\mathcal{S}_0]=f(\mathbf{v})$. $\square$

Two vectors $\mathbf{v}\prec\mathbf{z}$ are *dependent* with respect to $\prec$ if $\textsc{pr}[\phi^{-1}(\mathbf{v})|\mathbf{z}]>0$. Consider now a partial order $\prec'$ derived from $\prec$ by only retaining relations between dependent vectors. Then all linearizations of $\prec'$ have the same mapping of outcomes to determining vectors and thus, the resulting order-based estimators are identical. Conversely, when a *partial* order $\prec'$ has the property that for all outcomes $S$, $\min_{\prec}V^*(S)$ is unique, we can specify the estimator $f^{(\prec')}$ with respect to it (same as using any linearization).

LEMMA 3.2. *The estimator $\hat{f}^{(\prec)}$ is monotone if and only if for any outcome $S$ and $\mathbf{v}\in V^*(S)$, the estimate on outcomes determined by $\mathbf{v}$ is at least $\hat{f}^{(\prec)}(S)$:*

$$\hat{f}^{(\prec)}\text{ is monotone}\iff\forall S,\,\forall\mathbf{v}\in V^*(S)\;\hat{f}^{(\prec)}(\mathbf{v})\geq\hat{f}^{(\prec)}(S)\,.$$

PROOF. An outcome $S'$ with $V^*(S')=\{\mathbf{v}\}$ has $V^*(S')\subset V^*(S)$ and is determined by $\mathbf{v}$. From monotonicity, we must have $\hat{f}^{(\prec)}(\mathbf{v})\geq\hat{f}^{(\prec)}(S)$. Conversely, consider two outcomes $S$ and $S'$ such that $V^*(S)\subset V^*(S')$. Let $\mathbf{v}'$ be the determining vector of $S'$ and $\mathbf{v}$ be the determining vector of $S$. We have that $\mathbf{v}\in S'$, hence $\hat{f}^{(\prec)}(\mathbf{v})=\hat{f}^{(\prec)}(S)\geq\hat{f}^{(\prec)}(S')$. $\square$

**Forcing nonnegativity $\hat{f}^{(+\prec)}$:** When the constraints specifying $f^{(\prec)}$ have no nonnegative solution, we can explicitly constrain the setting of $\hat{f}^{(\prec)} : \mathcal{S}'$ to ensure that nonnegativity is not violated on *successive* vectors:

$$\min \sum_{S\in\mathcal{S}'} \textsc{pr}[S|\mathbf{v}](\hat{f}(S)-f(\mathbf{v}))^2 \quad (7)$$

$$\sum_{S\in\mathcal{S}'} \textsc{pr}[S|\mathbf{v}]\hat{f}(S) = f(\mathbf{v}) - f_0 \quad (8)$$

$$\forall \mathbf{v}' \succ \mathbf{v} \quad \sum_{S\in\mathcal{S}'\cup\mathcal{S}_0} \hat{f}(S)\textsc{pr}[S|\mathbf{v}'] \le f(\mathbf{v}') \,. \quad (9)$$

We minimize variance (7) subject to unbiasedness (8) and not violating nonnegativity to any $\mathbf{v}' \succ \mathbf{v}$ (9). The resulting estimator is Pareto optimal if the solution of the system is unique.

A solution $\hat{f}^{(+\prec)}$ satisfying nonnegativity constraints is identical to $\hat{f}^{(\prec)}$ when the latter is defined and nonnegative. With $\hat{f}^{(+\prec)}$ formulation, the constraints (9) can make two vectors $\mathbf{v}$ and $\mathbf{z}$ dependent also when both sets of outcomes $\phi^{-1}(\mathbf{v})$ and $\phi^{-1}(\mathbf{z})$ have positive probability for some data vector $\mathbf{y}$ that succeeds both $\mathbf{v}$ and $\mathbf{z}$. This is because when $\mathbf{v}$ precedes $\mathbf{z}$, the constraints (9) due to $\mathbf{y}$ are less tight. As with $\hat{f}^{(\prec)}$, we can equivalently define $\hat{f}^{(+\prec)}$ with respect to a partial order $\prec'$ derived from $\prec$ by including only relations between dependent data vectors and all linearizations of $\prec'$.

**Ordered partition $\hat{f}^{(\mathcal{U})}$:** The order-based formulations, however, in particular the more constrained $\hat{f}^{(+\prec)}$, can preclude symmetric estimators. Symmetric estimators are naturally desirable when $f$ is symmetric (invariant under permuting coordinates). When two symmetric vectors are dependent under $\prec$, the member that $\prec$-precedes the other can have a strictly lower variance.

We therefore seek a more relaxed formulation that will allow us to balance the variance of symmetric vectors.

We consider a setup where data vectors are partitioned into ordered batches $\mathcal{U} = \{U_0, U_1, \ldots\}$. The estimator $\hat{f}^{(\mathcal{U})}$ prioritizes earlier batches but "balances" the variance between vectors that are members of the same batch. That is, the estimator is *locally Pareto optimal* for each $U_i$: given $\hat{f} : \mathcal{S}_0$, unbiasedness (8) for all $\mathbf{v} \in U_i$ and nonnegativity (9) for all $\mathbf{v}' \in U_{>i}$. That is, under these constraints, there is no other setting of $\hat{f}$ on $\mathcal{S}'$ with smaller or equal variance for all vectors in $U_i$, and a strictly smaller variance for at least one vector. The estimator $\hat{f}^{(\mathcal{U})}$ is Pareto optimal if at each step $h$, when fixing the variance of all vectors in $U_h$, the solution is unique. Symmetry (invariance to permutation of entries) can be achieved by including all symmetric data vectors in the same part and using a symmetric locally optimal estimator.

This is formulated in Algorithm 2, which processes $U_i$ at step $i$, setting the estimator on all outcomes consistent with $U_i$ and not consistent with any vector in $U_j$ for $j < i$.

**Algorithm 2** $\hat{f}^{(\mathcal{U})}$

**Require:** $U_0, U_1, \ldots$ is a partition of $\mathbf{V}$
1: $\mathcal{S}_0 \leftarrow \emptyset$ ▷ set of processed outcomes
2: **for** $h = 0, 1, 2, \ldots$ **do** ▷ $h$ is the index of current part to process
3: $\mathcal{S}' \leftarrow \{S | U_h \cap V^*(S) \neq \emptyset\} \setminus \mathcal{S}_0$ ▷ Unprocessed outcomes consistent with $U_h$
4: Compute a locally optimal estimator for $U_h$, extending $\hat{f}$ on $\mathcal{S}_0$, and satisfying

$$\forall \mathbf{v}', \sum_{S\in\mathcal{S}'\cup\mathcal{S}_0} \hat{f}(S)\textsc{pr}[S|\mathbf{v}'] \le f(\mathbf{v}') \,.$$

5: $\mathcal{S}_0 \leftarrow \mathcal{S}_0 \cup \mathcal{S}'$

# 4. POISSON: WEIGHT-OBLIVIOUS

We now consider estimating $f(\mathbf{v})$ when sampling of entries is weight-oblivious and Poisson: entry $i \in [r]$ is sampled independently with probability $p_i > 0$.

The outcome $S \subset [r]$ includes the sampled entries, and for each sampled entry $i \in S$, the value $v_i$.

The inverse-probability estimate $\hat{f}^{(HT)}(S) = f(\mathbf{v})/\prod_{i\in[r]} p_i$, when $S \equiv [r]$ (all entries are sampled), and $\hat{f}^{(HT)}(S) = 0$ otherwise, is defined for all $f$ and from (1) has variance

$$\textsc{var}[\hat{f}^{(HT)}] = f(\mathbf{v})^2 \left(\frac{1}{\prod_{i\in[r]} p_i} - 1\right) \,. \quad (10)$$

This estimator is the optimal inverse probability estimator for quantiles and range: The set of outcomes $\mathcal{S}^*$ which contains all outcomes with $|S| = r$ is the most inclusive set for which we can determine both the value $f(\mathbf{v})$ and $\textsc{pr}[\mathcal{S}^*|\mathbf{v}]$ (see Section 2.2). The estimators $\hat{\textsc{rg}}^{(HT)}$ ($r = 2$) and $\hat{\min}^{(HT)}$ are even (Pareto) optimal: this is because any nonnegative estimator must have $\hat{f}(S) = 0$ on outcomes $v$ consistent with data vectors with $f(\mathbf{v}) = 0$, which includes all outcomes with $|S| < r$ for these two functions. Considering all estimators that assume positive values only when $|S| = r$, variance is minimized when using a fixed value. The estimator $\hat{f}^{(HT)}$, however, is not optimal for all other quantiles ($\ell^{\text{th}}$ when $\ell < r$) or for $\textsc{rg}$ when $r > 2$.

We present optimal estimators for max and Boolean OR: the monotone estimators $\hat{\max}^{(L)}$ and $\hat{\text{OR}}^{(L)}$ which prioritize dense data vectors and the estimators $\hat{\max}^{(U)}$ and $\hat{\text{OR}}^{(U)}$ which prioritize sparse vectors.

## 4.1 Estimator $\hat{\max}^{(L)}$

We compute the estimator $f^{(\prec)}$ (Algorithm 1) with respect to the following partial order $\prec$: The data vector $\mathbf{0}$ precedes all others, that is $\forall \mathbf{v} \in V$, $\mathbf{0} \prec \mathbf{v}$. Otherwise, $\prec$ corresponds to the numeric order on $L(\mathbf{v}) \equiv |\{j \in [r] \mid v_j < \max_{i\in[r]} v_i\}|$ (the number of entries strictly lower than the maximum one): $\mathbf{v} \prec \mathbf{w} \iff L(\mathbf{v}) < L(\mathbf{w})$.

For an outcome $S$, the set $V^*(S)$ includes all vectors that agree with the outcome on sampled entries: $\mathbf{v}' \in V^*(S) \iff \forall i \in S,\ v'_i = v_i$.

The determining vector $\phi(S)$ of an outcome $S$ is $\min_\prec V^*(S)$: $\phi(S) = \mathbf{0}$ if $\forall i \in S,\ v_i = 0$ (In particular if $S = \emptyset$). If $S \neq \emptyset$, $\phi(S)_j \equiv v_j$ if $j \in S$ and $\phi(S)_j = \max_{i\in S} v_i$ otherwise. The mapping $\phi(S)$ is well defined by $\prec$, which means that the estimator $f^{(\prec)}$ (if defined) is unique. Because $\prec$ is symmetric (invariant to permutation of entries), so is $f^{(\prec)}$.

Our choice of $\prec$ aims at obtaining a monotone estimator through conservative (low) estimate values: The determining vector of an outcome $S$ has all unsampled entries set to the maximum value of a sampled entry (this value is also the lower bound $\underline{f}(V^*(S'))$). The optimal estimate value for this vector on $S$ would be lower than if we had a determining vector with lower entries and same maximum because the outcome on such a vectors is more likely to have a lower maximum sampled entry, meaning a lower estimate on such outcomes which needs to be compensated for by a higher estimate on $S$.

For the minimum vector $\mathbf{0}$, there are no preceding outcomes ($\mathcal{S}_0 = \emptyset$) and we can directly compute $\hat{\max}^{(L)}(\mathbf{0})$ (the estimate for all outcomes that have $\phi(S) = \mathbf{0}$), obtaining $\hat{\max}^{(L)} = 0$ on all outcomes $S$ such that $\forall i \in S, v_i = 0$.

We can now proceed and compute the estimator for all outcomes $S$ with determining vector $\mathbf{v}$ such that $L(\mathbf{v}) = 0$, that is, outcomes where at least one entry is sampled, has positive value, and all other sampled entries have the same value: $\forall_{i\in S},\ v_i = \max_{i\in S} v_i > 0$. The probability of such an outcome given data vector $\mathbf{v}$ with $L(\mathbf{v}) = 0$ is the probability that at least one entry is sampled: $1 - \prod_{i\in[r]}(1-p_i)$ and the estimate value is accordingly

$$\hat{\max}^{(L)} = \frac{\max_{i\in S} v_i}{1-\prod_{i\in[r]}(1-p_i)} . \tag{11}$$

**Maximum over two instances ($r = 2$).** We have $\hat{\max}^{(L)} = 0$ on outcomes consistent with data $(0, 0)$ and from (11) $\hat{\max}^{(L)} = \frac{\max_{i\in S} v_i}{p_1+p_2-p_1p_2}$ for outcomes consistent with data with two equal positive entries ($S = \{1\}$, $S = \{2\}$, or $S = \{1, 2\}$ and $v_1 = v_2 = v$). We now consider data vectors where $v_2 < v_1$ (other case $v_1 < v_2$ is symmetric). The estimate is already computed on outcomes where exactly one entry is sampled. These and the empty outcome are in $\mathcal{S}_0$. The outcomes $\mathcal{S}'$ are those where both entries are sampled, and hence $\text{PR}[\mathcal{S}'] = p_1p_2$. To be unbiased, the estimate $x$ must satisfy the linear equation (line 13 of Algorithm 1):

$$\max\{v_1, v_2\} = p_1p_2x + \\ +p_1(1-p_2)\frac{v_1}{p_1+p_2-p_1p_2} + p_2(1-p_1)\frac{v_2}{p_1+p_2-p_1p_2} .$$

Solving and summarizing we obtain:

| Outcome $S$ | | $\hat{\max}^{(L)}(S)$ |
|---|---|---|
| $S=\emptyset$ | : | $0$ |
| $S=\{1\}$ | : | $\frac{v_1}{p_1+p_2-p_1p_2}$ |
| $S=\{2\}$ | : | $\frac{v_2}{p_1+p_2-p_1p_2}$ |
| $S=\{1,2\}$ | : | $\frac{\max(v_1,v_2)}{p_1p_2} - \frac{(1/p_2-1)v_1+(1/p_1-1)v_2}{p_1+p_2-p_1p_2}$ . |

Expressing the estimator as a function of the determining vector, assuming $v_1 \geq v_2$ (other case is symmetric), we obtain:

$$\hat{\max}^{(L)}(\mathbf{v}) = v_1\frac{1}{p_1(p_1+p_2-p_1p_2)} - v_2\frac{1-p_1}{p_1(p_1+p_2-p_1p_2)} \tag{12}$$

LEMMA 4.1. *The estimator* $\hat{\max}^{(L)}$ *is Pareto optimal, monotone, nonnegative, and dominates the estimator* $\hat{\max}^{(HT)}$.

PROOF. Pareto optimality follows from the $\hat{f}^{(\prec)}$ derivation. For monotonicity, we observe that determining vectors of more informative outcomes (outcomes with more entries sampled) have an equal-or-larger maximum entry $v_1$ or an equal-or-smaller minimum entry $v_2$, which clearly holds as the coefficient of $v_1$ in (12) is positive and that of $v_2$ is negative. Nonnegativity follows from monotonicity and the fact that the estimate is 0 when $S = \emptyset$.

The estimator $\hat{\max}^{(HT)}$ assumes values 0 or $\frac{\max(v_1,v_2)}{p_1p_2}$ and thus maximizes variance amongst all unbiased estimators with values in the same range. Hence, to establish dominance over $\hat{\max}^{(HT)}$, it suffices to show that on data $\mathbf{v}$, $\hat{\max}^{(L)}(v) \leq \frac{\max(v_1,v_2)}{p_1p_2}$, which is immediate from (12). □

**Multiple instances: $\hat{\max}^{(L)}$ for $r \geq 2$.**

A *sorting permutation* of a vector $\mathbf{v}$ is a permutation $\boldsymbol{\pi}$ of $[r]$ such that $v_{\pi_1} \geq \cdots \geq v_{\pi_r}$. We use the notation $\boldsymbol{\pi}(\boldsymbol{v}) = (v_{\pi_1}, \ldots, v_{\pi_r})$.

We prove that the estimator $\hat{\max}^{(L)}$ applied to an outcome $S$ can be expressed as a linear combination of the sorted entries of the determining vector $\phi(S)$. The coefficients depend on an accordingly permuted probability vector. When there are multiple entry of equal value, the sorting permutation is not unique. We show, however, that the estimator is invariant to the particular sorting permutation used.

THEOREM 4.1.

$$\hat{\max}^{(L)}(S) = \sum_{i\in[r]} \alpha_{i,\boldsymbol{\pi}(\boldsymbol{p})}\phi(S)_{\pi_i} , \tag{13}$$

*where $\boldsymbol{\pi}$ is the sorting permutation of $\phi(S)$ and $\alpha_{i,\boldsymbol{q}}$ are rational expressions in $q_1, \ldots, q_r$ that are always defined when $q_i \in (0, 1]$.*

*Moreover, the coefficients' prefix sums*

$$A_{h,\boldsymbol{p}} \equiv \sum_{i=1}^{h} \alpha_{i,\boldsymbol{p}} \tag{14}$$

*are* symmetric *rational expressions for $p_i$ for $i \in [h]$ and for $p_i$ for $i \in [r] \setminus [h]$.*

PROOF. We first show that the symmetry property of the prefix sums implies that the estimate does not depend on the choice of sorting permutation (when it is not unique). It suffices to show this for a sorted $\mathbf{v}$ such that $v_j \equiv v_{j+1}$ and show that when symmetry holds, the estimator is the same for the identity permutation $(1, \ldots, r)$ and the permutation $(1, \ldots, j-1, j+1, j, j+1, \ldots, r)$ (exchanging positions $j$ and $j + 1$). Both are sorting permutations of $\mathbf{v}$). The argument can be applied repeatedly if there are more than two equal entries.

Let $\mathbf{v}$ be sorted and let $\delta_i \equiv v_{i+1} - v_i$ for $i =\in [r-1]$. We can rewrite (13) as

$$\sum_{i=1}^{r} \alpha_{i,\boldsymbol{p}}v_i = A_{r,\boldsymbol{p}}v_1 - \sum_{i=1}^{r-1}\delta_i A_{i,\boldsymbol{p}}$$

When $\delta_j = 0$, let $\boldsymbol{p}$ and $\boldsymbol{p'}$ respectively be the original and permuted vectors with $p_j$ and $p_{j+1}$ exchanged. By symmetry, $A_{i,\boldsymbol{p}} = A_{i,\boldsymbol{p'}}$ for $i \in [r] \setminus \{j\}$. But $\delta_j = 0$, and hence the estimator is the same with both permutations.

We now show that the estimator has the form (13) and that the prefix sums satisfy the symmetry property. For $\mathbf{v}$ with sorting permutation $\boldsymbol{\pi}$ and $L(\mathbf{v}) = k$, we can rewrite (13) as

$$\sum_{i=1}^{h} \alpha_{i,\boldsymbol{\pi}(\boldsymbol{p})}v_{\pi_i} \tag{15}$$

$$= v_{\pi_1}A_{r-k,\boldsymbol{\pi}(\boldsymbol{p})} + \sum_{i=r-k+1}^{r}(A_{i,\boldsymbol{\pi}(\boldsymbol{p})} - A_{i-1,\boldsymbol{\pi}(\boldsymbol{p})})v_{\pi_i} .$$

For all outcomes $S$ consistent with $\mathbf{v}$, $L(\phi(S)) \leq L(\mathbf{v}) \leq k$. Thus, the estimator for data $\mathbf{v}$ is fully specified by $A_{h,\boldsymbol{\pi}(\boldsymbol{p})}$ where $h \geq r - L(\mathbf{v})$.

We show by induction on $k \geq 0$, that the estimator can be expressed in this form for data vectors with $L(\mathbf{v}) \leq k$. For the base case of the induction ($k = 0$), it suffices to specify the rational expression $A_{r,\boldsymbol{p}}$. By substituting a determining vector with all entries equal in (13) and equating with (11), we obtain

$$A_{r,\boldsymbol{p}} = \frac{1}{1-\prod_{i\in[r]}(1-p_i)} \tag{16}$$

specifies $\max^{(L)}(\mathbf{v})$ for all determining vectors $\mathbf{v}$ such that $L(\mathbf{v}) = 0$ (all entries are equal and positive) and thus specifies the estimator correctly for all data vectors with $L(\mathbf{v}) = 0$. Symmetry clearly holds as $A_{r,\boldsymbol{\pi}(\boldsymbol{p})}$ is independent of the particular permutation $\boldsymbol{\pi}$.

In the induction step we assume that the rational expressions $A_{i,\boldsymbol{p}}$ are well defined and satisfy symmetry for all $i \geq r-k$ and all $\boldsymbol{p}$, that is, (15) is equal to $\max^{(L)}$ when $L(\phi(S)) \leq k$ and hence the estimator is specified for data with $L(\mathbf{v}) \leq k$. We then specify $A_{r-k-1,\boldsymbol{p}}$ by relating it through a linear equation to higher prefix sums. This fully specifies the estimator for data with $L(\mathbf{v}) = k+1$. Symmetry properties of $A_{r-k-1,\boldsymbol{\pi}(\boldsymbol{p})}$ (showing it is symmetric in $\{p_1, \ldots, p_{r-k-1}\}$ and in $\{p_{r-k}, \ldots, p_r\}$) follow from the symmetry in the equation and assumed symmetry of the higher prefix sums.

We now express $A_{r-k-1,\boldsymbol{p}}$ as a linear combination of prefix sums of the form $A_{h,\boldsymbol{\pi}'(\boldsymbol{p})}$ where $h \geq r-k$ and $[r-k] \subset \{\pi'_1, \ldots, \pi'_h\}$.

Consider a vector $\mathbf{z}$ such that $L(\mathbf{z}) = k+1$ and entries are sorted in nonincreasing order (the sorting permutation is the identity, and this is without loss of generality as we can permute $\boldsymbol{p}$ accordingly). We show that there is a (unique) value of $A_{r-k-1,\boldsymbol{p}}$ that results in an unbiased estimate for $\mathbf{z}$. This value turns out to be independent of $\mathbf{z}$ (works for all vectors with $L(\mathbf{v}) = k+1$ and same permutation of sorted entries). When solved parametrically, this is a rational expression in $p_1, \ldots, p_r$ that satisfies the symmetry property.

The vector $\mathbf{z}$ has $z_1 = \cdots = z_{r-k-1} \equiv Z$ and $z_r \leq \cdots \leq z_{r-k} < Z$. Consider the vector $\mathbf{z}'$ that is equal to $\mathbf{z}$ on all entries except that $z'_{r-k} \equiv Z$. Clearly $L(\mathbf{z}') = k$ and therefore, by induction, the estimate for $\mathbf{z}'$ is unbiased, that is, has expectation $Z$ on data $\mathbf{z}'$.

We relate outcomes for different data vectors that correspond to the same *sample* $\boldsymbol{\sigma} \in 2^{[r]}$ which is the set of sampled entries. The vectors $\mathbf{z}'$ and $\mathbf{z}$ have the same determining vectors, and thus, the same estimate on all samples where $\sigma_{r-k} = 0$ (do not include the entry $r-k$). Therefore, the estimate is unbiased on $\mathbf{z}$, if and only if the expectation for data $\mathbf{z}$ is equal to the expectation for data $\mathbf{z}'$ over samples where $\sigma_{r-k} = 1$ (entry $r-k$ is sampled).

We consider the difference in the contribution to the estimate of a sample $\boldsymbol{\sigma}$ that includes $r-k$ on the data vectors $\mathbf{z}$ and $\mathbf{z}'$.

If none of the entries $[r-k-1]$ is sampled, the determining vectors differ on the first $h \geq r-k$ entries (where $h$ is equal to $r-k$ plus the number of unsampled entries in $[r-k+1, r]$). The value of the determining vector on the first $h$ entries is $Z$ when the data is $\mathbf{z}$ and $Z'$ when the data is $\mathbf{z}'$. There is a sorting permutation $\boldsymbol{\pi}'$ for both determining vectors which depends only on the sample $\boldsymbol{\sigma}$ (works for all choices of $\mathbf{z}$ and respective $\mathbf{z}'$: it has all unsampled entries in sorted order, followed by entry $r-k$, and then by other sampled entries in sorted order. Thus, the difference in the contribution to the estimate is $A_{h,\boldsymbol{\pi}'(\boldsymbol{p})}(Z - Z')$.

If at least one of the entries $[r-k-1]$ is sampled, then the determining vectors are identical on the first $h-1$ entries (value is $Z$), differ on entry $h$ (the value is $Z$ when data is $\mathbf{z}$ and $Z'$ when $\mathbf{z}'$) and identical on remaining entries (values smaller than $Z'$). Again, there is a common sorting permutation $\boldsymbol{\pi}'$ for the determining vector of all choices of $\mathbf{z}$ and of $\mathbf{z}'$: it contains the first $r-k-1$ entries and unsampled entries in $[r-k+1, r]$, all in sorted order, followed by $r-k$, and then sampled entries in $[r-k+1, r]$ in sorted order (note that it is the same permutation we used for the case where none of the entries $[r-k-1]$ are sampled). Thus, the difference in the contribution to the estimate is $\alpha_{h,\boldsymbol{\pi}'(\boldsymbol{p})}(Z - Z') = (A_{h,\boldsymbol{\pi}'(\boldsymbol{p})} - A_{h-1,\boldsymbol{\pi}'(\boldsymbol{p})})$(Z-Z'). The only samples for which $h = r-k$ is when all entries in $[r-k, r]$ are sampled. In this case the determining vectors are the respective data vectors and the sorting permutation of the determining vector is the identity. Thus the only "unknown" is $A_{r-k-1,\boldsymbol{p}}$ and it appears, when replacing $\alpha_{r-k,\boldsymbol{p}} = A_{r-k,\boldsymbol{p}} - A_{r-k-1,\boldsymbol{p}}$.

Recall that for the estimate for $\mathbf{z}$ to be unbiased, the expectation of these differences over samples must be 0. The expectation is the sum over samples $\boldsymbol{\sigma}$ of the probability of the sample

$$\text{PR}[\boldsymbol{\sigma}] = \prod_{i \in [r]} p_i^{\sigma_i}(1-p_i)^{1-\sigma_i}$$

multiplied by the difference. By equating with 0 we obtain a linear equation with one variable $A_{r-k-1,\boldsymbol{p}}$, which must have a unique solution. Since all terms are multiplied by $(Z - Z')$, it factors out. The equation and solution $A_{r-k-1,\boldsymbol{p}}$ are independent of $\mathbf{z}$. Therefore, the estimate is unbiased for all data vectors $\mathbf{z}$ with $L(\mathbf{z}) = k+1$.

□

We now write the equations explicitly, using the notation

$$\begin{aligned} \boldsymbol{\pi}^{\boldsymbol{\sigma}} &= (1, \ldots, r-k-1, \{i = r-k+1, \ldots, r | \sigma_i = 0\}, \\ & \quad r-k, \{i = r-k+1, \ldots, r | \sigma_i = 1\}) \\ h^{\boldsymbol{\sigma}} &= r-k + \sum_{i=r-k+1}^{r} \sigma_i \end{aligned}$$

for the sorting permutation and $h$ value used with the sample $\boldsymbol{\sigma}$.

$$\begin{aligned} 0 &= \sum_{\boldsymbol{\sigma} \in 2^{[r]} | \sigma_{r-k}=1} \text{PR}[\boldsymbol{\sigma}] \Big( I(\sum_{i=1}^{r-k-1} \sigma_i = 0) A_{h^{\boldsymbol{\sigma}},\boldsymbol{\pi}^{\boldsymbol{\sigma}}(\boldsymbol{p})} + \\ & \quad I(\sum_{i=1}^{r-k-1} \sigma_i \geq 1) \alpha_{h^{\boldsymbol{\sigma}},\boldsymbol{\pi}^{\boldsymbol{\sigma}}(\boldsymbol{p})} \Big) \end{aligned}$$

where $I$ is the indicator function.

We can express the equation in terms of the projection $\overline{\boldsymbol{\sigma}}$ of the sample $\boldsymbol{\sigma}$ on entries $K = [r-k+1, r]$. We combine terms with identical projection while noting that $\boldsymbol{\pi}^{\boldsymbol{\sigma}} \equiv \boldsymbol{\pi}^{\overline{\boldsymbol{\sigma}}}$ and $h^{\boldsymbol{\sigma}} \equiv h^{\overline{\boldsymbol{\sigma}}}$ depend only on the projection. We eliminate common terms and obtain:

$$\begin{aligned} 0 &= \sum_{\overline{\boldsymbol{\sigma}} \in 2^K} \text{PR}[\overline{\boldsymbol{\sigma}}] \Big( \prod_{i=1}^{r-k-1} (1-p_i) A_{h^{\overline{\boldsymbol{\sigma}}},\boldsymbol{\pi}^{\overline{\boldsymbol{\sigma}}}(\boldsymbol{p})} + \\ & \quad (1 - \prod_{i=1}^{r-k-1} (1-p_i))(A_{h^{\overline{\boldsymbol{\sigma}}},\boldsymbol{\pi}^{\overline{\boldsymbol{\sigma}}}(\boldsymbol{p})} - A_{h^{\overline{\boldsymbol{\sigma}}}-1,\boldsymbol{\pi}^{\overline{\boldsymbol{\sigma}}}(\boldsymbol{p})}) \Big) \\ &= \sum_{\overline{\boldsymbol{\sigma}} \in 2^K} \text{PR}[\overline{\boldsymbol{\sigma}}] \qquad (17) \\ & \quad \Big( A_{h^{\overline{\boldsymbol{\sigma}}},\boldsymbol{\pi}^{\overline{\boldsymbol{\sigma}}}(\boldsymbol{p})} - (1 - \prod_{i=1}^{r-k-1} (1-p_i)) A_{h^{\overline{\boldsymbol{\sigma}}}-1,\boldsymbol{\pi}^{\overline{\boldsymbol{\sigma}}}(\boldsymbol{p})} \Big) \end{aligned}$$

For $k = 0$, $K = \emptyset$ and thus there is only one term. The equation relates $A_{r-1,\boldsymbol{p}}$ and $A_{r,\boldsymbol{p}}$, yielding

$$A_{r-1,\boldsymbol{p}} = \frac{A_{r,\boldsymbol{p}}}{(1 - \prod_{i=1}^{r-1}(1-p_i))} \qquad (18)$$

For $k = 1$, $K = \{r\}$ and hence there are two terms, according to the value of $\sigma_r$:

$$\begin{aligned} 0 &= (1-p_r)\Big(A_{r,\boldsymbol{p}} - \\ &\quad (1-\prod_{i=1}^{r-2}(1-p_i))A_{r-1,(p_1,\ldots,p_{r-2},p_r,p_{r-1})}\Big) \\ &\quad +p_r\Big(A_{r-1,\boldsymbol{p}} - (1-\prod_{i=1}^{r-2}(1-p_i))A_{r-2,\boldsymbol{p}}\Big) \end{aligned}$$

Therefore,

$$A_{r-2,\boldsymbol{p}} = \frac{A_{r-1,\boldsymbol{p}} + A_{r-1,\boldsymbol{p'}} - A_{r,\boldsymbol{p}}}{1-\prod_{i=1}^{r-2}(1-p_i)}$$

where $\boldsymbol{p'} = (p_1,\ldots,p_{r-2},p_r,p_{r-1})$.

We conjecture that $\hat{\max}^{(L)}$ is monotone, nonnegative, and dominates $\hat{\max}^{(HT)}$. We verified these properties for $r \leq 4$ with uniform $\boldsymbol{p}$, using the following lemma and explicit computation of the coefficients.

LEMMA 4.2. *To establish monotonicity, nonnegativity, and dominance of* $\hat{\max}^{(L)}$ *over* $\hat{\max}^{(HT)}$ *it suffices to show that* $\alpha_i < 0$ *for* $i > 1$ *and that* $\alpha_1 \leq 1/\prod_{i\in[r]} p_i$.

PROOF. To establish monotonicity, consider two types of manipulations of a determining vector: increasing some of its maximum entries or decreasing a maximum entry in case the maximum entry is not unique. Now, for any data $\mathbf{v}$ and outcomes $S_1 \subset S_2$ ($S_2$ contains all entries sampled in $S_1$ and more), the determining vector of $S_2$ can be obtained from that of $S_1$ using such operations. For monotonicity, we need to show that the estimate value obtained for $\mathbf{v}$ on outcome $S_2$ is at least that of $S_1$, equivalently, that these manipulations can only increase $\hat{\max}^{(L)}$. For the second manipulation, it suffices to show that $\alpha_i < 0$ for $i > 1$. For the first manipulation, it suffices to show that $\sum_{j=1}^{i} \alpha_j > 0$ for all $i \geq 1$. Since we know that $\sum_{i\in[r]} \alpha_i > 0$, this is implied by $\alpha_i < 0$ for $i > 1$. Nonnegativity follows from monotonicity and the base case of estimate value 0 when there are no sampled entries.

To establish dominance over $\hat{\max}^{(HT)}$, given monotonicity, it suffices to show that $\alpha_1 \leq 1/\prod_{i\in[r]} p_i$. This means that all $\hat{\max}^{(L)}$ estimates on a given data vector $\mathbf{v}$ are at most $\frac{\max(\mathbf{v})}{\prod_{i\in[r]} p_i}$, which is the $\hat{\max}^{(HT)}$ estimate. The HT estimate has maximum variance amongst all unbiased estimators that assume values in the range $\left[0, \frac{\max(\mathbf{v})}{\prod_{i\in[r]} p_i}\right]$. Hence, $\mathrm{VAR}[\hat{\max}^{(L)}] \leq \mathrm{VAR}[\hat{\max}^{(HT)}]$. □

These expression can be use to compute the estimator, but the number of different prefix-sums grows exponentially with the number of distinct probabilities in the $k$ suffix of $\boldsymbol{p}$. We give specific consideration to uniform probabilities.

**Uniform $\boldsymbol{p}$.** When $p = p_1 = p_2 = \ldots = p_r$, we can use $\alpha_{i,\boldsymbol{p}} \equiv \alpha_{i,p}$ and $A_{i,\boldsymbol{p}} \equiv A_{i,p}$ for the coefficients in (13) and their respective prefix sums. For a given $p$, we only need $r$ different values, $A_{i,p}$ for $i \in [r]$, to specify the estimator. We omit $p$ from the subscript for brevity. We show that for a fixed $p$, the estimator can be computed in time quadratic in the dimension.

THEOREM 4.2. *The estimator, for a given $p$, can be computed in $O(r^2)$ time using the relation*

$$A_{r,p} = \frac{1}{1-(1-p)^r} \tag{19}$$

$$A_{r-k-1} = \tag{20}$$
$$\frac{A_{r-k} + \sum_{\ell=1}^{k}\binom{k}{\ell}\left(\frac{1-p}{p}\right)^{\ell}\left(A_{r-k+\ell} - (1-(1-p)^{r-k-1})A_{r-k+\ell-1}\right)}{1-(1-p)^{r-k-1}}$$

PROOF. Using uniform $p$, (16) simplifies to (19). The equation (17) simplifies to

$$\begin{aligned} 0 &= \sum_{\ell=0}^{k}\binom{k}{\ell}p^{k-\ell}(1-p)^{\ell}\cdot \\ &\quad \Big(A_{r-k+\ell} - (1-(1-p)^{r-k-1})A_{r-k+\ell-1}\Big) \end{aligned} \tag{21}$$

We obtain (20) by expressing $A_{r-k-1}$ as a function of $A_h$ for $h \geq r-k$. This relation is a triangular system of linear equations and allows us to compute the estimator (the coefficients $\alpha_{i,p}$ for $i \in [r]$ for a given $p$) in time $O(r^2)$. □

We compute the parametric form of the higher prefix sums:

$$\begin{aligned} A_r &= \frac{1}{1-(1-p)^r} \\ A_{r-1} &= A_r\frac{1}{1-(1-p)^{r-1}} \\ A_{r-2} &= A_{r-1}\frac{1+(1-p)^{r-1}}{1-(1-p)^{r-2}} \end{aligned}$$

For $r = 2$, we obtain

$$\begin{aligned} A_2 &= \frac{1}{p(2-p)} \\ A_1 &= \frac{1}{p^2(2-p)} \end{aligned}$$

Using $\alpha_2 = A_2 - A_1$ and $\alpha_1 = A_1$, we obtain the estimator

$$\boldsymbol{\alpha} = \left(\frac{1}{p^2(2-p)}, -\frac{1-p}{p^2(2-p)}\right) \tag{22}$$

For $r = 3$, we obtain

$$\begin{aligned} A_3 &= \frac{1}{p(p^2-3p+3)} \\ A_2 &= \frac{1}{p^2(p^2-3p+3)(2-p)} \\ A_1 &= \frac{2+p^2-2p}{p^3(p^2-3p+3)(2-p)} \end{aligned}$$

Using $\alpha_3 = A_3 - A_2$, $\alpha_2 = A_2 - A_1$, and $\alpha_1 = A_1$, the estimator is

$$\begin{aligned} \boldsymbol{\alpha} &= \Big(\frac{2-2p+p^2}{p^3(2-p)(3-3p+p^2)}, \\ &\quad -\frac{1-p}{p^3(3-3p+p^2)}, -\frac{(1-p)^2}{p^2(2-p)(3-3p+p^2)}\Big) \end{aligned}$$

Algorithm 3 includes pseudo-code for the computation of the coefficients and for the application of the estimator $\hat{\max}^{(L)}$ for uniform $p$ and any $r > 1$.

**Algorithm 3** $\hat{\max}^{(L)}$ uniform $p$

```
function COEFF(r,p)                         ▷ compute coefficients of estimator
    A_r ← 1/(1−(1−p)^r)                     ▷ prefix sums
    for k = 0,1,2,...,r−2 do
        t ← Σ_{ℓ=1}^{k} (k choose ℓ) ((1−p)/p)^ℓ (A_{r−k+ℓ} − (1 − (1 − p)^{r−k−1}) A_{r−k+ℓ−1})
        A_{r−k−1} ← (A_{r−k} + t)/(1−(1−p)^{r−k−1})

    α_1 ← A_1                               ▷ compute coefficients
    for h = 2,...,r do
        α_h ← A_h − A_{h−1}
    return α

function EST(S, α)                          ▷ Estimator applied to outcome S
    if S = ∅ then return 0
    z ← SORTDEC{v_i | i ∈ S}                ▷ multiset of values of
      sampled entries is sorted in non-increasing order
                                            ▷ Compute sorted determining vector u
    for i = 1,...,|S| do
        u_{i+r−|S|} ← z_i
    for i = 1,...,r − |S| do
        u_i ← z_1
    return Σ_{i=1}^{r} α_i u_i
```

## 4.2 Estimator $\hat{\max}^{(U)}$

We now seek an estimator which prioritizes "sparse" vectors, which is captured by order-optimal where vectors with fewer positive entries precede others. Formally, we use an ordered partition according to $L(\mathbf{v}) \equiv |\{j \in [r] | v_j > 0\}|$, where part $U_h$ includes all vectors with $L(\mathbf{v}) = h$. We derive estimators for $r = 2$, while demonstrating usage of the different constructions.

The minimum vectors are $U_0 \equiv \mathbf{0}$. An outcome $S$ is consistent with $\mathbf{0}$ if and only if $\forall i \in S, v_i = 0$ and we set $\hat{\max}^{(U)}(S) \leftarrow 0$. This setting must be the same for all nonnegative unbiased estimators.

We first attempt to apply Algorithm 1. The determining vector $\phi(S)$ is uniquely defined by the partial order $\prec$ and obtained by substituting 0 for all unsampled entries $i \notin S$. This, the estimator is invariant to a choice of a total order linearizing $\prec$. Processing $U_1$, we obtain the estimate $\hat{\max}^{(U)}(S) = v_i/p_i$ on all outcomes with one positive entry $v_i > 0$ amongst $i \in S$. It remains to process vectors $U_2$. The outcomes $\mathcal{S}'$ have $S = \{1,2\}$ with $v_1, v_2 > 0$ and hence a determining vector with two positive entries. The estimate is the solution of the linear equation $p_1p_2\hat{\max}^{(U)}(S) + p_1(1-p_2)\frac{v_1}{p_1} + p_2(1-p_1)\frac{v_2}{p_2} = \max((v_1, v_2))$. The solution, $\frac{\max(v_1,v_2)-(1-p_1)v_2-(1-p_2)v_1}{p_1p_2}$, however, may be negative (e.g., when $v_1 = v_2$ and $p_1 + p_2 < 1$).

To obtain a nonnegative $\prec$-optimal estimator, we must enforce the nonnegativity constraints (9) when processing $U_1$. Now the result is sensitive to the particular order of processing vectors in $U_1$: Suppose vectors of the form $(v_1, 0)$ are processed before vectors of the form $(0, v_2)$. The vector $(v_1, 0)$ is the determining vector of all outcomes with the first entry sampled. That is, all outcomes with both entry sampled and values are $(v_1, v_2)$ and outcomes with only the first entry sampled and has value $v_1$. The probability of such outcome given data $(v_1, 0)$ is $p_1$. To minimize variance, we would like to set the estimate to $v_1/p_1$ on these outcomes, which we can do because this setting does not violate nonnegativity (9) for other vectors. We next process vectors of the form $(0, v_2)$. They are determining vectors for outcomes $\mathcal{S}'_1$ with both entries sampled and values are $(0, v_2)$ and outcomes $\mathcal{S}'_2$ with only the second entry sampled and value is $v_2$. The outcomes $\mathcal{S}'_1$ are not consistent with any other data, and are not constrained by (9). The outcomes $\mathcal{S}'_2$ are also consistent with data vectors with two positive entries $(v'_1, v_2)$ and therefore we need to ensure that we do not violate (9) for these vectors. To minimize the variance on $(0, v_2)$, we seek $\hat{\max}(\mathcal{S}_1) \geq \hat{\max}(\mathcal{S}_2)$ with $\hat{\max}(\mathcal{S}_2)$ being as large as possible without violating (9). Lastly, we process vectors with two positive entries. The outcomes determined by these vectors have both entries sampled and are not consistent with any other data vector. Summarizing, we obtain the estimator

| Outcome $S$ | | $\hat{\max}^{(Uas)}(S)$ |
|---|---|---|
| $S = \emptyset$ | : | 0 |
| $S = \{1\}$ | : | $\frac{v_1}{p_1}$ |
| $S = \{2\}$ | : | $\frac{v_2}{\max\{1-p_1, p_2\}}$ |
| $S = \{1,2\}$ | : | $\frac{\max(v_1,v_2) - \frac{p_2(1-p_1)}{\max\{1-p_1,p_2\}}v_2 - (1-p_2)v_1}{p_1p_2}$ |

This estimator is Pareto optimal but is asymmetric: the estimate changes if the entries of $\mathbf{v}$ (and $\boldsymbol{p}$) are permuted.

To obtain a symmetric estimator, we apply Algorithm 2 processing $U_1$ and $U_2$ in batches, searching for a symmetric locally optimal estimator for $U_1$ and then for $U_2$. We obtain:

| Outcome $S$ | | $\hat{\max}^{(U)}(S)$ |
|---|---|---|
| $S = \emptyset$ | : | 0 |
| $S = \{1\}$ | : | $\frac{v_1}{p_1(1+\max\{0,1-p_1-p_2\})}$ |
| $S = \{2\}$ | : | $\frac{v_2}{p_2(1+\max\{0,1-p_1-p_2\})}$ |
| $S = \{1,2\}$ | : | $\frac{\max(v_1,v_2) - \frac{v_1(1-p_2)+v_2(1-p_1)}{1+\max\{0,1-p_1-p_2\}}}{p_1p_2}$ |

We can see that $\hat{\max}^{(U)}$ dominates $\hat{\max}^{(HT)}$ – this follows from $\hat{\max}^{(U)} \leq \max(\mathbf{v})/(p_1p_2)$ for data $(v_1, v_2)$.

**Example.** Figure 1 illustrates the relation between $\hat{\max}^{(L)}$, $\hat{\max}^{(U)}$, and $\hat{\max}^{(HT)}$ and their variance when data vectors have the form $\mathbf{v} = (v_1, v_2)$ and each entry is sampled independently with probability 1/2. The plot shows the ratios $\frac{\text{VAR}[\hat{\max}^{(L)}]}{\text{VAR}[\hat{\max}^{(HT)}]}$ and $\frac{\hat{\max}^{(U)}}{\text{VAR}[\hat{\max}^{(HT)}]}$ as a function of $\min(v_1, v_2)/\max(v_1, v_2)$. We can see that $\hat{\max}^{(HT)}$ is dominated by $\hat{\max}^{(L)}$ and $\hat{\max}^{(U)}$ and that the two Pareto optimal estimators $\hat{\max}^{(L)}$ and $\hat{\max}^{(U)}$ are incomparable: on inputs where one of the values is 0, $\text{VAR}[\hat{\max}^{(U)}] = \frac{3}{4}\max(\mathbf{v})^2$ whereas $\text{VAR}[\hat{\max}^{(L)}] = (11/9)\max(\mathbf{v})^2$. On inputs where $v_1 = v_2$, $\text{VAR}[\hat{\max}^{(L)}] = (1/3)\max(\mathbf{v})^2$ whereas $\text{VAR}[\hat{\max}^{(U)}] = \frac{3}{4}\max(\mathbf{v})^2$.

## 4.3 Boolean OR

We now consider $\text{OR}(\mathbf{v}) = v_1 \vee v_2 \vee \cdots \vee v_r$ over the domain $\mathbf{V} = \{0,1\}^r$. The best inverse probability estimator is $\hat{\text{OR}}^{(HT)} = 1/\prod_{i=1}^{r} p_i$ when $|S| = r$ and $\bigvee_{i \in S} v_i = 1$ and $\hat{\text{OR}}^{(HT)} = 0$ otherwise. By specializing $\hat{\max}^{(L)}$ and $\hat{\max}^{(U)}$, we obtain the estimators $\hat{\text{OR}}^{(L)}$ and $\hat{\text{OR}}^{(U)}$, which turn out to be optimal also in this more restricted domain. Optimality of $\hat{\text{OR}}^{(L)}$ follows from order optimality with respect to $\prec$ satisfying: $\forall \mathbf{v} \in V \setminus \{\mathbf{0}\}, \mathbf{0} \prec \mathbf{v}$

Sample distribution:

| | $1 \in S$ | $1 \notin S$ |
|---|---|---|
| $2 \notin S$ | $1/4$ | $1/4$ |
| $2 \in S$ | $1/4$ | $1/4$ |

| $\hat{\max}^{(HT)}$ | $1 \in S$ | $1 \notin S$ |
|---|---|---|
| $2 \notin S$ | $0$ | $0$ |
| $2 \in S$ | $4\max(v_1, v_2)$ | $0$ |

| $\hat{\max}^{(L)}$ | $1 \in S$ | $1 \notin S$ |
|---|---|---|
| $2 \notin S$ | $\frac{4v_1}{3}$ | $0$ |
| $2 \in S$ | $\frac{8\max(v_1,v_2)-4\min(v_1,v_2)}{3}$ | $\frac{4v_2}{3}$ |

| $\hat{\max}^{(U)}$ | $1 \in S$ | $1 \notin S$ |
|---|---|---|
| $2 \notin S$ | $2v_1$ | $0$ |
| $2 \in S$ | $2\max(v_1, v_2) - 2\min(v_1, v_2)$ | $2v_2$ |

$\textsc{var}[\hat{\max}^{(HT)}] = 3\max(v_1, v_2)^2$

$$\begin{aligned}\textsc{var}[\hat{\max}^{(L)}] &= \frac{11}{9}\max(v_1, v_2)^2 + \frac{8}{9}\min(v_1, v_2)^2 - \frac{16}{9}\max(v_1, v_2)\min(v_1, v_2) \\ &\leq \frac{11}{9}\max(v_1, v_2)^2\end{aligned}$$

$$\begin{aligned}\textsc{var}[\hat{\max}^{(U)}] &= \frac{3}{4}\max(v_1, v_2)^2 + 2\min(v_1, v_2)^2 - 2\max(v_1, v_2)\min(v_1, v_2) \\ &\leq \frac{3}{4}\max(v_1, v_2)^2\end{aligned}$$

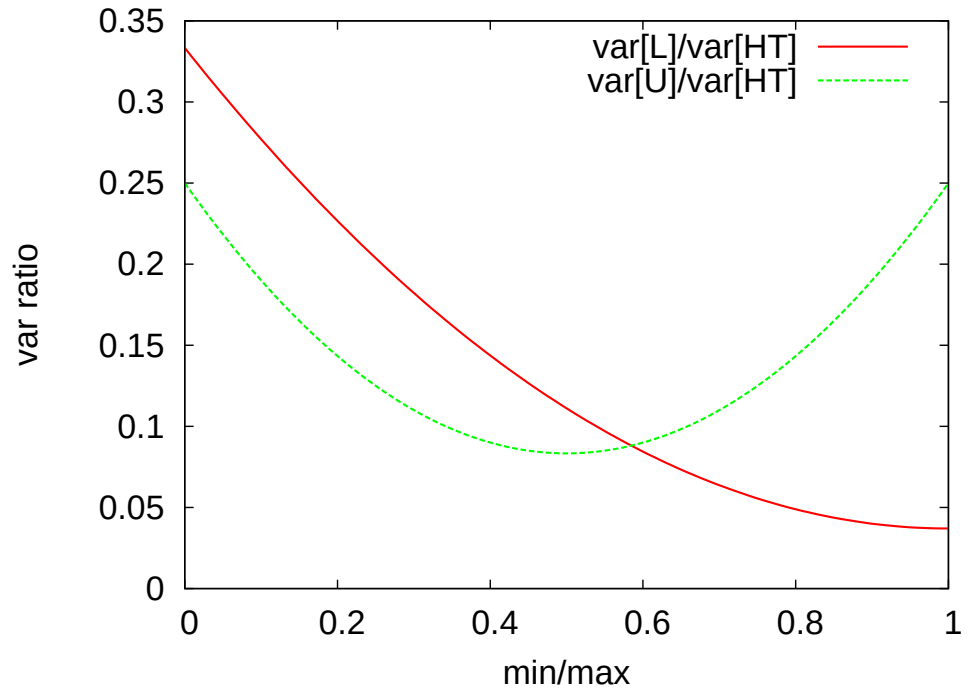

**Figure 1: Estimators for $\max\{v_1, v_2\}$ over Poisson samples (weight-oblivious) with $p_1 = p_2 = 1/2$.**

and for $\mathbf{v}, \mathbf{v}' \neq \mathbf{0}$,

$$\mathbf{v} \prec \mathbf{v}' \iff L(\mathbf{v}) < L(\mathbf{v}') \ ,$$

where $L(\mathbf{v}) = |\{i | v_i = 0\}|$ is the number of zero entries in $\mathbf{v}$. The determining vector $\phi(S)$ is obtained by setting, for $i \notin S$, $v_i \leftarrow \bigvee_{j \in S} v_j$. For $r = 2$, the estimator as a function of the determining vector is

$$\hat{\mathrm{OR}}^{(L)}(v_1, v_2) = \frac{\mathrm{OR}(v_1, v_2)}{p_1 p_2} - \frac{(1/p_2 - 1)v_1 + (1/p_1 - 1)v_2}{p_1 + p_2 - p_1 p_2} \ .$$

Optimality of $\hat{\mathrm{OR}}^{(U)}$ follows by noticing that when specializing the construction of $\hat{\max}^{(U)}$, the construction remains optimal with respect to an ordered partition according to $r - L(\mathbf{v})$.

**Variance.** To gain a better understanding of the relative performance of the estimators $\hat{\mathrm{OR}}^{(HT)}$, $\hat{\mathrm{OR}}^{(L)}$, and $\hat{\mathrm{OR}}^{(U)}$, we study their variance. For data $\mathbf{0}$, all estimates are 0, and thus all three estimators have zero variance, On all data $\mathbf{v}$ with $\mathrm{OR}(\mathbf{v}) = 1$, using (1):

$$\textsc{var}[\hat{\mathrm{OR}}^{(HT)} | \mathrm{OR}(\mathbf{v}) = 1] = \frac{1}{\prod_{i \in [r]} p_i} - 1 \ . \quad (23)$$

The variance of $\hat{\mathrm{OR}}^{(L)}$ and $\hat{\mathrm{OR}}^{(U)}$, has more fine dependence on the data vector: The estimate $\hat{\mathrm{OR}}^{(L)}$ on data vector $(1, 1)$ is $1/p$ with probability $p = p_1 + p_2 - p_1 p_2$ and 0 otherwise and hence, using (1):

$$\textsc{var}[\hat{\mathrm{OR}}^{(L)} | (1, 1)] = \frac{1}{p_1 + p_2 - p_1 p_2} - 1 \ . \quad (24)$$

The estimate for data vector $(1, 0)$ is 0 with probability $1 - p_1$ (entry 1 is not sampled), $\frac{1}{p_1+p_2-p_1p_2}$ with probability $p_1(1 - p_2)$ ($S = \{1\}$) and $\frac{1}{p_1(p_1+p_2-p_1p_2)}$ when $S = \{1, 2\}$. Therefore,

$$\begin{aligned}&\textsc{var}[\hat{\mathrm{OR}}^{(L)} | (1, 0)] \\ &= (1 - p_1) + p_1(1 - p_2)(\frac{1}{p_1 + p_2 - p_1 p_2} - 1)^2 \\ &+ p_1 p_2 (\frac{1}{p_1(p_1 + p_2 - p_1 p_2)} - 1)^2\end{aligned}$$

Figure 2 shows the variance of the estimators $\hat{\mathrm{OR}}^{(HT)}$, $\hat{\mathrm{OR}}^{(L)}$ and $\hat{\mathrm{OR}}^{(U)}$ as a function of $p = p_1 = p_2$. The estimators $\hat{\mathrm{OR}}^{(L)}$ and $\hat{\mathrm{OR}}^{(U)}$ dominate $\hat{\mathrm{OR}}^{(HT)}$. The estimator $\hat{\mathrm{OR}}^{(L)}$ has minimum variance on $(1, 1)$ and $\hat{\mathrm{OR}}^{(U)}$ is the symmetric estimator with minimum variance on $(1, 0)$ and $(0, 1)$ (over all nonnegative unbiased estimators).

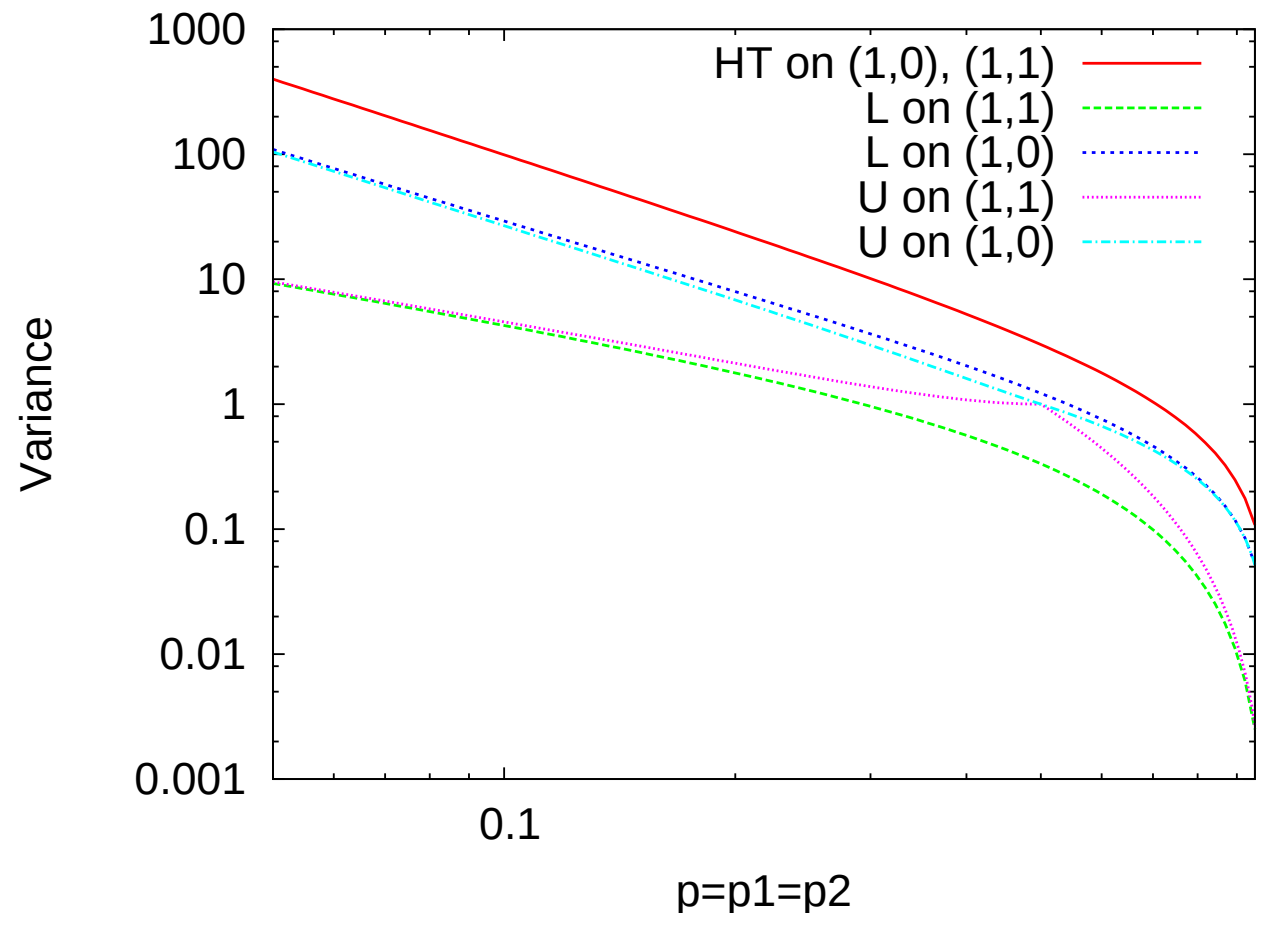

**Figure 2: Variance of $\hat{\mathrm{OR}}^{(HT)}$, $\hat{\mathrm{OR}}^{(L)}$ and $\hat{\mathrm{OR}}^{(U)}$ when $p_1 = p_2 = p$, on data vectors $(1, 1)$ and $(1, 0)$, as a function of $p$.**

Asymptotically, when $p \to 0$, for data vectors $(1,0)$, $(0,1)$, and $(1,1)$ we get that $\textsc{var}[\hat{\text{OR}}^{(HT)}] \approx 1/p^2$. On the other hand, for the data vectors $(1,0)$ and $(0,1)$ we have that $\textsc{var}[\hat{\text{OR}}^{(L)}]$, $\textsc{var}[\hat{\text{OR}}^{(U)}] \approx 1/(4p^2)$ and for the data vector $(1,1)$, $\textsc{var}[\hat{\text{OR}}^{(L)}]$, $\textsc{var}[\hat{\text{OR}}^{(U)}] \approx 1/(2p)$.

This means that for data $(1,1)$ ("no change"), the variance is half the square root of the variance of $\hat{\text{OR}}^{(HT)}$. For data $(1,0)$ or $(0,1)$ ("change") variance is $1/4$ of the variance of $\hat{\text{OR}}^{(HT)}$.

# 5. POISSON: WEIGHTED, KNOWN SEEDS

We turn our attention to weighted Poisson sampling with known seeds, starting with estimating OR over binary domains and then consider max over the nonnegative reals.

For the purpose of deriving estimators over *binary domains* ($\mathbf{V} = \{0,1\}^r$), Poisson weighted sampling with known seeds is equivalent to Poisson weight-oblivious sampling (Section 4). This relation holds only for binary domains and is established through a 1-1 mapping between outcomes in terms of the information we can glean from the outcome.

The sample distribution of weighted sampling over binary domains is as follows: there is a seed vector $\mathbf{u} \in [0,1]^r$ where $u_i$ are independent and selected uniformly at random from the interval $[0,1]$. Defining $\boldsymbol{p} \in [0,1]^r$ such that $p_i = \textsc{pr}[\tau_i \leq 1]$,

$$i \in S \iff v_i = 1 \wedge u_i \leq p_i \ .$$

$p_i$ is the probability that the $i$th entry is sampled if $v_i = 1$. The entry is never sampled if $v_i = 0$ but since we know $\mathbf{u}$, if $u_i \leq p_i$ and $i \notin S$ we know that $v_i = 0$.

We now map an outcome $S$ of weighted sampling with known seeds to outcome $S'$ of weight-oblivious sampling with vector $\boldsymbol{p}$

$$\begin{array}{rcl} i \in S & \iff & i \in S' \text{ and } v_i = 1 \\ i \notin S \text{ and } u_i \leq p_i & \iff & i \in S' \text{ and } v_i = 0 \\ i \notin S \text{ and } u_i > p_i & \iff & i \notin S' \ . \end{array}$$

It is easy to see that $\textsc{pr}[S] = \textsc{pr}[S']$ and that $V^*(S) \equiv V^*(S')$.

Observe that the weighted sample $S$ is smaller than the corresponding weight-oblivious one $S'$ since entries with 0 values are not represented in the sample. Knowledge of seeds, however, compensates for this. We use knowledge of the seeds in a more elaborate way in the (significantly more involved) derivations of estimators for $\max(\mathbf{v})$.

## 5.1 Boolean OR

We state the estimators $\hat{\text{OR}}^{(HT)}$, $\hat{\text{OR}}^{(L)}$, and $\hat{\text{OR}}^{(U)}$ by mapping the respective estimators obtained in the weight-oblivious setting (Section 4.3).

The optimal inverse-probability estimator uses the set of outcomes $\mathcal{S}^*$ such that $\forall i \in [r], u_i \leq p_i$. This corresponds to $S = [r]$ in the weight-oblivious setting. If $\forall i \in [r], u_i \leq p_i$ and $\text{OR}(\mathbf{v}) = 1$, $\hat{\text{OR}}^{(HT)} = 1/\prod_{i\in[r]} p_i$. Otherwise, $\hat{\text{OR}}^{(HT)} = 0$.

**Estimator $\hat{\text{OR}}^{(L)}$.**

| Outcome $S$ | | $\hat{\text{OR}}^{(L)}$ |
|---|---|---|
| $S = \emptyset$ | : | 0 |
| $(S = \{1\} \wedge u_2 > p_2) \vee$ | | |
| $(S = \{2\} \wedge u_1 > p_1) \vee$ | | |
| $S = \{1,2\}$ | : | $\frac{1}{p_1+p_2-p_1p_2}$ |
| $S = \{1\} \wedge u_2 \leq p_2$ | : | $\frac{1}{p_1(p_1+p_2-p_1p_2)}$ |
| $S = \{2\} \wedge u_1 \leq p_1$ | : | $\frac{1}{p_2(p_1+p_2-p_1p_2)}$ |

**Estimator $\hat{\text{OR}}^{(U)}$.**

| Outcome $S$ | | $\hat{\text{OR}}^{(U)}$ |
|---|---|---|
| $S = \emptyset$ | : | 0 |
| $S = \{1\} \wedge u_2 > p_2$ | : | $\frac{1}{p_1(1+\max\{0,1-p_1-p_2\})}$ |
| $S = \{2\} \wedge u_1 > p_1$ | : | $\frac{1}{p_2(1+\max\{0,1-p_1-p_2\})}$ |
| Else | : | $\frac{1-\frac{v_1(1-p_2)+v_2(1-p_1)}{1+\max\{0,1-p_1-p_2\}}}{p_1p_2}$ |

The variance of the estimators is the same as in the weight oblivious case (see Section 4.3 and Figure 2). In Section 8.1 we show how our $\hat{\text{OR}}$ estimators can be applied to estimate distinct element count (union of sets), which are the sum aggregates of OR.

## 5.2 Maximum over nonnegative reals

We study estimating max under Poisson PPS weighted sampling. The seed vector $\mathbf{u} \in [0,1]^r$ has entries drawn independently and uniformly from $[0,1]$. $\boldsymbol{\tau}^*$ is a fixed vector and an entry $i$ is included in $S$ iff $v_i \geq u_i\tau_i^*$, that is, with probability $\min\{1, v_i/\tau_i^*\}$.

Recall that both $\boldsymbol{\tau}^*$ and the seed vector $\mathbf{u}$ are available to the estimator. Therefore, when $i \notin S$, we know that $v_i < u_i\tau_i^*$.

# Estimator $\hat{\max}^{(HT)}$ [17, 18]

Consider the set of outcomes $\mathcal{S}^*$ such that

$$S \in \mathcal{S}^* \iff \max_{i \notin S} u_i\tau_i^* \leq \max_{i\in S} v_i \ .$$

This set includes all outcomes $S$ from which $\max(\mathbf{v})$ can be determined: For $S \in \mathcal{S}^*$, $\max(\mathbf{v}) = \max_{i\in S} v_i$. For any data vector $\mathbf{v}$, the probability that the outcome is in $\mathcal{S}^*$

$$\textsc{pr}[\mathcal{S}^* \mid \mathbf{v}] = \prod_{i\in[r]} \min\{1, \max_{i\in S} v_i/\tau_i^*\} \ ,$$

can be computed from the outcome, for any outcome in $\mathcal{S}^*$. The inverse-probability estimator is therefore:

$$\hat{\max}^{(HT)}(S) = \begin{cases} \text{if } \max_{i\notin S} u_i\tau_i^* \leq \max_{i\in S} v_i & : \dfrac{\max_{i\in S} v_i}{\prod_{i\in[r]} \min\left\{1, \max_{i\in S} v_i/\tau_i^*\right\}} \\ \text{otherwise} & : 0 \end{cases}$$

This is the optimal inverse-probability estimator since $\mathcal{S}^*$ is the most inclusive set possible.

# Estimator $\hat{\max}^{(L)}$

We use the partial order $\prec$ with $\mathbf{0}$ preceding all other vectors, and otherwise the order corresponds to an increasing lexicographic order on the lists $L(\mathbf{v})$ that is the sorted multiset of differences $\{\max(\mathbf{v}) - v_i \mid i \in [r]\}$.

$\hat{\max}^{(L)}$ is ordered-based with respect to $\prec$ and is defined through Algorithm 1. For an outcome $S$, the set $V^*(S)$ of consistent vectors contains all vectors with $v_i$ as in $S$ for $i \in S$ and $v_i \leq u_i\tau_i^*$ otherwise. The minimum consistent data vector $\min_\prec V^*(S)$ is well defined and thus $\phi(S) = \min_\prec V^*(S)$ is $\mathbf{0}$ when $S = \emptyset$ and otherwise has $\phi(S)_i = v_i$ for $i \in S$ and $\phi(S)_i = \min\{\max_{j\in S} v_j, u_i\tau_i^*\}$ for $i \notin S$. Note that when $S \neq \emptyset$, all entries of $\phi(S)$ are positive.

The estimator $\hat{\max}^{(L)}$ for $r = 2$ is presented in Figure 3 using two tables. The first table shows a mapping of outcomes to determining vectors, the second states the estimator as a function of the determining vector. The derivation is in Appendix A. Monotonicity, nonnegativity, and bounded variance can be easily verified for $r = 2$ and are conjectured for $r > 2$.

| outcome $S$ | | determining vector $\phi(S)$ | |
|---|---|---|---|
| | | $\phi(S)_1$ | $\phi(S)_2$ |
| $S=\emptyset$ | : | 0 | 0 |
| $S=\{1\}$ | : | $v_1$ | $\min\{u_2\tau_2^*, v_1\}$ |
| $S=\{2\}$ | : | $\min\{u_1\tau_1^*, v_2\}$ | $v_2$ |
| $S=\{1,2\}$ | : | $v_1$ | $v_2$ |

| $\mathbf{v}=(v_1,v_2)$, $v_1\geq v_2$ | | $\hat{\max}^{(L)}(\mathbf{v})$ |
|---|---|---|
| $\mathbf{v}=(0,0)$ | : | 0 |
| $v_1\geq v_2\geq \tau_2^*$ | : | $v_2+\frac{v_1-v_2}{\min\{1,\frac{v_1}{\tau_1^*}\}}$ |
| $v_1\geq \tau_1^*$, $v_2\leq \min\{\tau_2^*, v_1\}$ | : | $v_1$ |
| $v_2\leq v_1\leq \min\{\tau_1^*,\tau_2^*\}$ | : | $\frac{\tau_1^*\tau_2^*}{\tau_1^*+\tau_2^*-v_1}+\frac{\tau_1^*\tau_2^*(\tau_1^*-v_1)}{v_1(\tau_1^*+\tau_2^*)}\ln\left(\frac{(\tau_1^*+\tau_2^*-v_2)v_1}{v_2(\tau_1^*+\tau_2^*-v_1)}\right)+\frac{(v_1-v_2)\tau_1^*\tau_2^*(\tau_1^*-v_1)}{v_1(\tau_1^*+\tau_2^*-v_2)(\tau_1^*+\tau_2^*-v_1)}$ |
| $v_2\leq \tau_2^*\leq v_1\leq \tau_1^*$ | : | $\tau_1^*+\tau_2^*-\frac{\tau_1^*\tau_2^*}{v_1}+\frac{(\tau_1^*\tau_2^*)(\tau_1^*-v_1)}{v_1(\tau_1^*+\tau_2^*)}\ln\left(\frac{(\tau_1^*+\tau_2^*-v_2)\tau_1^*}{\tau_2^*(\tau_1^*+\tau_2^*-v_1)}\right)+\frac{\tau_2^*(\tau_1^*-v_1)(\tau_2^*-v_2)}{(\tau_1^*+\tau_2^*-v_2)v_1}$ |

**Figure 3: Estimator $\hat{\max}^{(L)}$ for $r=2$. The top table maps each outcome $S$ to the determining vector $\phi(S)$. The bottom table presents the estimator as a function of the determining vector $\mathbf{v}$ when $v_1\geq v_2$ (symmetric expressions for the case $v_2\geq v_1$ are omitted).**

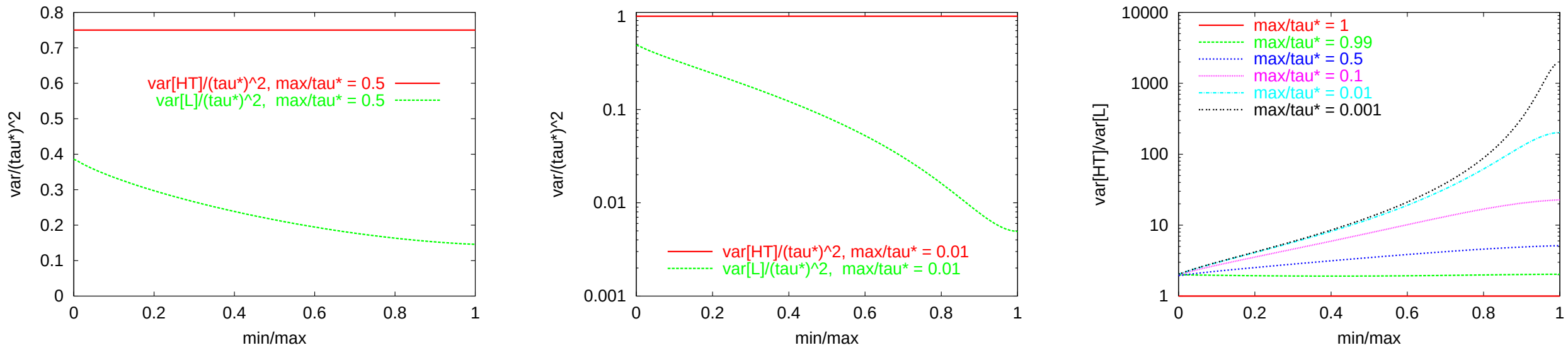

(A) (B) (C)

**Figure 4: Estimators $\hat{\max}^{(L)}$ and $\hat{\max}^{(HT)}$ for two independent pps samples with $\tau_1^*=\tau_2^*=\tau^*$. (A) and (B) show the normalized variance $\textsc{var}[\hat{\max}]/(\tau^*)^2$ for $\rho=\max(v_1,v_2)/\tau^*\in\{0.01,0.5\}$, as a function of $\min(v_1,v_2)/\max(v_1,v_2)$. (C) shows the variance ratio $\textsc{var}[\hat{\max}^{(HT)}]/\textsc{var}[\hat{\max}^{(L)}]$ as a function of $\min(v_1,v_2)/\max(v_1,v_2)$ for different values of $\rho$.**

**Variance.** Figure 4 illustrates the relation between $\textsc{var}[\hat{\max}^{(L)}]$ and $\textsc{var}[\hat{\max}^{(HT)}]$ when $\tau_1^*=\tau_2^*=\tau^*$. The estimator $\hat{\max}^{(L)}$ dominates $\hat{\max}^{(HT)}$. We show the variance (divided by $(\tau^*)^2$ ) as a function of the ratio $\min(\mathbf{v})/\max(\mathbf{v})$. When $\max(\mathbf{v})\geq\tau^*$ or $\mathbf{v}=\mathbf{0}$, $\textsc{var}[\hat{\max}^{(HT)}\mid\mathbf{v}]=\textsc{var}[\hat{\max}^{(L)}\mid\mathbf{v}]=0$, and these are the only cases where there is no advantage to $\hat{\max}^{(L)}$ over $\hat{\max}^{(HT)}$. For all other data vectors,

$$\frac{\textsc{var}[\hat{\max}^{(HT)}|\mathbf{v}]}{\textsc{var}[\hat{\max}^{(L)}|\mathbf{v}]}\geq\frac{1+\rho}{\rho}\geq 2\,,$$

where $\rho=\max(\mathbf{v})/\tau^*$. That is the variance ratio is at least 2 and asymptotically $O(1/\rho)$ when $\rho$ is small.

Fixing $\rho$, the inverse-probability weight estimator is positive with probability $p=(\frac{\max(\mathbf{v})}{\tau^*})^2=\rho^2$. Hence, $\frac{\textsc{var}[\hat{\max}^{(HT)}|\mathbf{v}]}{(\tau^*)^2}=\rho^2(1/p-1)=1-\rho^2$ and is independent of $\min(\mathbf{v})$. The variance of the $\hat{\max}^{(L)}$ estimator decreases with $\min(\mathbf{v})$. For a fixed $\rho$, it is minimized when $\min(\mathbf{v})=\max(\mathbf{v})$ and is maximized when $\min(\mathbf{v})=0$ ($\mathbf{v}=(\rho\tau^*,0)$ or $\mathbf{v}=(0,\rho\tau^*)$). For the vector $\mathbf{v}=(0,\rho\tau^*)$ the $\hat{\max}^{(L)}$ estimator equals $\tau^*=\rho\tau^*/\rho$ with probability $\rho$ and 0 otherwise so

$$\frac{\textsc{var}[\hat{\max}^{(L)}|(\rho\tau^*,0)]}{(\tau^*)^2}=\rho^2(1/\rho-1)=\rho-\rho^2\,.$$

The variance ratio is accordingly at least

$$\frac{\textsc{var}[\hat{\max}^{(HT)}]}{\textsc{var}[\hat{\max}^{(L)}]}\geq\frac{1-\rho^2}{\rho-\rho^2}=\frac{1+\rho}{\rho}\,.$$

The variance ratio $\textsc{var}[\hat{\max}^{(HT)}]/\textsc{var}[\hat{\max}^{(L)}]$ is larger when entry values are closer and with higher sampling rates (larger $\tau^*$). In Section 8.2 we apply $\hat{\max}^{(L)}$ to estimate the max dominance norm, which is the sum aggregate of max.

**Exponentiated Range:** There is no inverse-probability weight estimate for $\textsc{rg}_d$ ($d>0$), because on data vectors with $\min(\mathbf{v})=0$ there is 0 probability of determining $\textsc{rg}(\mathbf{v})$ from the outcome. We derive order-based optimal estimators for $\textsc{rg}_d$ ($d>0$) in [16].

# 6. POISSON: WEIGHTED, UNKNOWN SEEDS

We show that when seeds are not available to the estimator, it is not possible to obtain a nonnegative unbiased estimator for $\ell^{\text{th}}(\mathbf{v})$ where $\ell<r$ and for $\textsc{rg}_d$ ($d>0$) with weighted Poisson sampling. This impossibility results also holds for Boolean values and estimating OR and XOR of 2 or more bits.

This result is related to a negative result by Charikar et. al [7] for estimating distinct counts, which is the sum aggregate of the OR primitive. They showed that most of the data set needs to be sampled in order to obtain a constant error in constant probability on the distinct count. Their model essentially corresponds to sampling with unknown seeds.

This result completes our understanding of when nonnegative unbiased quantile estimators over Poisson samples exist: Inverse-probability weight estimators exist when sampling is weight-oblivious (Section 4), when weighted and seeds are known ([17, 18] and Section 5) and when weighted with unknown seeds for estimating min ($\ell = r$) (we obtain inverse-probability weights with respect to $\mathcal{S}^*$ that includes all outcomes with $S = [r]$).

THEOREM 6.1. *For any $\ell < r$, there is no unbiased nonnegative estimator for $\ell^{\text{th}}(\mathbf{v})$ over independent weighted samples with unknown seeds.*

PROOF. Recall that with weighted sampling, an entry where $v_i = 0$ is never sampled. As seeds are not available, we do not have any information from the outcome on values of entries that are not sampled. Therefore, the set $V^*(S)$ of data vectors consistent with $S$ includes all vectors in $\mathbf{V}$ that agree with $S$ on sampled entries.

We first establish the claim for $r = 2$. Since our arguments use values restricted to $\{0, 1\}$, they also hold for $\text{OR}(v_1, v_2)$. Let $p_i$ be the inclusion probability of entry $i$ when $v_i = 1$. We show that when $p_1 + p_2 < 1$, there is no unbiased estimator that is simultaneously correct for the four data vectors $(1, 1), (1, 0), (0, 1), (0, 0)$.

On outcome $S = \emptyset$, we must have $\hat{\text{OR}}(S) \equiv 0$ to ensure nonnegative estimates on data $(0, 0)$. When $S = \{i\}$ $(v_i = 1)$ the estimator must have expected value $1/p_i$ in order to be unbiased for $(1, 0)$ or $(0, 1)$. When the data is $(1, 1)$, the contribution to the expectation from outcomes with exactly one sampled entry is $p_1(1-p_2)/p_1 + p_2(1-p_1)/p_2 = 2 - p_1 - p_2 > 1$. In order to be unbiased, the estimator must have negative expectation on outcome $S = \{1, 2\}$, which contradicts nonnegativity.

Lastly, we extend the argument for $\ell^{\text{th}}(\mathbf{v})$ and general $r$. We consider the four data vectors where $v_3 = \cdots = v_{\ell+1} = 1, v_{\ell+2} = \cdots = v_r = 0$, and $(v_1, v_2) \in \{0, 1\}^2$. Let $p_i > 0$ be the sampling probability of entry $i$ when $v_i = 1$ and assume that $p_1 + p_2 < 1$. On these vectors, $\ell^{\text{th}}(\mathbf{v}) = \text{OR}(v_1, v_2)$. If neither 1 or 2 are sampled, we have $\ell - 1$ positive sampled entries and the estimate must be 0. On outcomes with exactly one $i \in \{1, 2\}$ sampled, the expectation of the estimator must be $\frac{1}{p_i \prod_{h=3}^{\ell+1} p_h}$ to be unbiased for data vectors $(v_1, v_2) = (1, 0), (0, 1)$. The contribution of the estimator from these outcomes for data with $v_1 = v_2 = 1$ is $\frac{2-p_1-p_2}{\prod_{h=3}^{\ell+1} p_h} > 1$, a contradiction. $\square$

The argument for $\text{RG}_d$ $(d > 0)$ is simpler. Consider estimating XOR of two bits with possible data $(0, 0)$, $(1, 1)$, and $(1, 0)$. The estimate value must be zero on outcomes with only one sampled entry. This is needed to guarantee nonnegativity for data vectors where the other unseen entry is equal to the sampled one. Consider now data $(1, 0)$. The two possible outcomes are that only the first entry is sampled or that neither entry is sampled with zero estimate value in both cases. Thus, the expectation of the estimator is 0 whereas $\text{RG}_d(1, 0) = 1$. A contradiction to unbiasedness.

# 7. ESTIMATING SUM AGGREGATES

When data is aggressively sampled, our basic estimators for individual quantile or range query have high variance. When the query is an aggregate – the sum of many basic queries, we can estimate it through the sum of the respective basic estimators. Since our estimators are unbiased, when estimates are independent, variance is additive and the relative error decreases with aggregation.

The data is modeled as a set $I$ of *instances*, where each instance $i \in I$ is an assignment of values (weights) to a set of keys $K$. For a key $h$, $\mathbf{v}(h)$ is the vector containing the values of $h$ in different instances. That is, entry $i$ of this vector, $v_i(h)$, is the value of key $h \in K$ in instance $i \in I$. Figure 5 (A) shows a data set with 3 instances $I = \{1, 2, 3\}$ and 6 keys $K = \{1, \ldots, 6\}$.

*Sum aggregates* have the form $\sum_{h \in K'} f(\mathbf{v}(h))$, where $K' \subset K$ are selected keys. The primitives (functions $f$) include *quantiles* (max, min, $\ell^{\text{th}}$ largest entry) and exponentiated range $\text{RG}_d = (\max(\mathbf{v}) - \min(\mathbf{v}))^d$ and are applied to values of a single-key across multiple instances. The sum aggregates for max, min, and RG over two instances are known as the *max-dominance norm*, *min-dominance norm* [19, 20], and $L_1$ distance. The $L_2$ distance is the square-root of a sum-aggregate of $\text{RG}_2$. When values are binary, each instance can be viewed as a set, and the sum aggregate for OR is the number of distinct keys (or the size of the union).

For the example data set in Figure 5(A), the max dominance norm over even keys ($K' = \{2, 4, 6\}$) and instances $\{1, 2\}$ is $10 + 20 + 10 = 40$. The $L_1$ distance between instances $\{2, 3\}$ over keys $K' = \{1, 2, 3\}$ is $10 + 5 + 3 = 18$.

**Applications**

Primary data sources structured as instances of values assigned to keys are *snapshots* of a changing data such as terms and their frequencies or sensor locations and measurement values and *request logs* recording activity (values) for different resources (keys): number of requests to each URL in Web traffic logs and bytes sent to each destination IP addresses in network traffic logs.

We classify queries as *single-instance*, *multi-instance*, or *decomposable*. Single-instance queries are over data from a single instance and decomposable queries can be stated as a nonnegative sum of single-instance queries, and can be estimated using a corresponding sum of single-instance estimators. Multi-instance queries are those that involve multiple instances and can not be decomposed, and are the ones targeted in our work. A single-instance query example on daily request logs is "total number of requests to .gov URLs on Monday." A decomposable query example is "total number of requests to .gov URLs in the past week," which can be posed as the sum of single-instance queries. Multi-instance queries include difference norms and distinct counts across days

We aim to find an optimal estimator when the query and underlying sampling scheme are given. The choice of sampling scheme is according to efficiency of processing on the data source and effectiveness on the queries of interest. Since the same sample might be used for different classes of queries, it is not necessarily optimized for a particular one. We review sampling methods, starting with single-instance, and the joint distributions (coordinated or independent) for multiple instances, and then show how to estimate sum aggregates using single-key estimators.

## 7.1 Sampling a single instance

We review popular summarization methods of a single instance: Poisson, bottom-$k$, and VAROPT sampling.

Sampling can be *weighted* or *weight-oblivious*. If sampling is weighted then the probability of including each key in the sample depends on its value $v(h)$. When sampling is weight-oblivious then inclusion probability does not depend on value.

The distinction between weighted and weight-oblivious sampling is important, even when values are binary, when sampling sparse data sets involving multiple keys. When most keys have zero values and only positive values are explicitly represented in the data, a weighted sampling algorithm needs to process only these keys and generates a sample containing only such keys whereas weight-oblivious sampling is applied to the full domain of keys (example is the set of active destination IP addresses at a given gateway, which is a small fraction of the key space of all possible IP addresses.)

Bottom-$k$ and Poisson samples are defined through a *random*

**keys:** $1, \ldots, 6$
**Instances:** $1, 2, 3$

| Instance/key | 1 | 2 | 3 | 4 | 5 | 6 |
|---|---|---|---|---|---|---|
| 1 | 15 | 0 | 10 | 5 | 10 | 10 |
| 2 | 20 | 10 | 12 | 20 | 0 | 10 |
| 3 | 10 | 15 | 15 | 0 | 15 | 10 |

**Example functions $f$**

| | | | | | | |
|---|---|---|---|---|---|---|
| $\max(v_1, v_2)$ | 20 | 10 | 12 | 20 | 10 | 10 |
| $\max(v_1, v_2, v_3)$ | 20 | 15 | 15 | 20 | 15 | 10 |
| $\min(v_1, v_2)$ | 15 | 0 | 10 | 0 | 0 | 10 |
| $\textsc{rg}(v_1, v_2, v_3)$ | 10 | 15 | 5 | 20 | 15 | 0 |

(A)

**Consistent shared-seed PPS ranks:**

| key: | 1 | 2 | 3 | 4 | 5 | 6 |
|---|---|---|---|---|---|---|
| $u$ | 0.22 | 0.75 | 0.07 | 0.92 | 0.55 | 0.37 |
| $r_1$ | 0.0147 | $+\infty$ | 0.007 | 0.184 | 0.055 | 0.037 |
| $r_2$ | 0.011 | 0.075 | 0.0583 | 0.046 | $+\infty$ | 0.037 |
| $r_3$ | 0.022 | 0.05 | 0.0047 | $+\infty$ | 0.0367 | 0.037 |

**Independent PPS ranks:**

| key: | 1 | 2 | 3 | 4 | 5 | 6 |
|---|---|---|---|---|---|---|
| $u_1$ | 0.22 | 0.75 | 0.07 | 0.92 | 0.55 | 0.37 |
| $r_1$ | 0.0147 | $+\infty$ | 0.007 | 0.184 | 0.055 | 0.037 |
| $u_2$ | 0.47 | 0.58 | 0.71 | 0.84 | 0.25 | 0.32 |
| $r_2$ | 0.0235 | 0.058 | 0.0592 | 0.042 | $+\infty$ | 0.032 |
| $u_3$ | 0.63 | 0.92 | 0.08 | 0.59 | 0.32 | 0.80 |
| $r_3$ | 0.063 | 0.0613 | 0.0053 | $+\infty$ | 0.0213 | 0.08 |

(B)

bottom-3 samples (shared seed):
1 : 3, 1, 6
2 : 1, 6, 4
3 : 3, 1, 5

bottom-3 samples (independent):
1 : 3, 1, 6
2 : 1, 6, 4
3 : 3, 5, 2

(C)

**Figure 5: (A): Example data set with keys $K = \{1, \ldots, 6\}$ and instances $\{1, 2, 3\}$. (B): per-key values for example aggregates. (C): random rank assignments and corresponding 3-order samples.**

*rank assignment* [36, 9, 12, 22, 13, 14] $r$ which maps keys to ranks. Rank values of different keys are independent. For each key $h$, the dependence of the rank value $r(h)$ on the weight $v(h)$ is captured by a family of probability density functions $\mathbf{f}_w$ ($w \geq 0$): The rank $r(h)$ is drawn from $\mathbf{f}_{v(h)}$.

- A *Poisson* sample is specified by a threshold $\tau$ or an expected sample size $k$ where $k = \sum_h \mathbf{F}_{v(h)}(\tau)$ and $\mathbf{F}_w$ is the CDF of $\mathbf{f}_w$. The sample is the set of keys with $r(h) < \tau$. Since ranks of different keys are independent, so are inclusions in the sample.

- A *bottom-$k$* sample contains the $k$ keys of smallest rank.

We can decouple the dependency of the rank on $v(h)$ from its dependency on the randomization: Each key obtains (independently) a *random seed* value $u(h) \in U[0, 1]$. The rank is then determined by the seed $u(h)$ and the value $v(h)$ to be $r(h) \leftarrow \mathbf{F}^{-1}_{v(h)}(u(h))$. Two families $\mathbf{f}_w$ that are used for weighted sampling.

- EXP ranks: $\mathbf{f}_w(x) = we^{-wx}$ ($\mathbf{F}_w(x) = 1 - e^{-wx}$) are exponentially-distributed with parameter $w$ (denoted by EXP$[w]$). Equivalently, if $u \in U[0, 1]$ then $-\ln(u)/w$ is an exponential random variable with parameter $w$. EXP$[w]$ ranks have the useful property that the minimum rank over a subpopulation $K'$ has distribution EXP$[v(K')]$, where $v(K') = \sum_{h \in K'} v(h)$. A bottom-$k$ sample is equivalent to taking $k$ weighted samples without replacement, where at each step a key is selected with probability equal to the ratio of $v(h)$ and the total value of the remaining keys [35, 28, 36, 9, 11, 23, 12, 13].

- PPS ranks: $\mathbf{f}_w$ is the uniform distribution $U[0, 1/w]$ ($\mathbf{F}_w(x) = \min\{1, wx\}$). This is the equivalent to choosing rank value $u/w$, where $u \in U[0, 1]$. The Poisson-$\tau$ sample is a PPS sample [28] (Inclusion Probability Proportional to Size). The bottom-$k$ sample is a priority sample [33, 22] (PRI).

Poisson sampling has the disadvantage that actual sample size varies. Bottom-$k$ sampling has fixed sample size but the dependence between keys complicates the design of the estimators and their analysis. VarOpt samples [10, 6], which we do not define here, have PPS inclusion probabilities and a fixed sample size. In VarOpt samples inclusion probabilities of different keys have nonpositive correlations which improves estimation quality. It is not clear, however, if we can incorporate "known seeds" into VarOpt sampling.

Bottom-$k$, Poisson, and VarOpt sampling are efficiently implemented on a data stream. Poisson sampling, where inclusions of different keys are independent, is applicable even when sampling of different keys must be completely decoupled (such as with transmitting sampled sensor measurements).

**Estimators**

We estimate sum aggregates using linear estimators of the form $\sum_{h \in K'} \hat{f}(h)$. An estimate $\hat{f}(h)$ is assigned to each key such that positive estimates are assigned only to keys included in the sample and estimates of other keys are 0. It follows that the estimate of the sum aggregate over $K'$ is equal to the sum of the individual estimates of keys included in the sample $S$: $\sum_{h \in K' \cap S} \hat{f}(h)$. From linearity of expectation, when the estimates $\hat{f}(h)$ are unbiased, so is the estimate of the sum.

The HT estimator, which assigns inverse-probability weights to sampled keys, is applicable to Poisson and VarOpt samples, where inclusion probabilities are available. With bottom-$k$ samples, the inclusion probability of a key depends on the weight distribution of all other keys. Tight unbiased estimators for subpopulation queries over bottom-$k$ samples were proposed only recently [22, 38, 12, 13]. The main insight was a delicate application of HT: the estimate for each key was obtained by applying inverse-probability weighting under the conditioning that the rank values of all other keys were fixed. This method, which we termed *rank conditioning* (RC) facilitated treating bottom-$k$ samples like independent samples for the purposes of estimator design. While clearly, conditioning increases variance with respect to the unattainable HT estimates, it turns out that performance loss is minimal [38].

## 7.2 Multiple Instances

Dispersed multiple instances are summarized independently, and therefore, for each key, the sampling of each entry $v_i(h)$ of the data vector $\mathbf{v}(h)$ is independent of the values of other entries. Samples of different instances can be independent, which is a model we studied here in more detail, but can also be coordinated.

**Coordinated sampling:** Estimation of many multi-instance functions, including quantile and difference queries, can be significantly improved by coordinating the sampling of different instances. Coordination means that a key that is sampled in one instance is more likely to be sampled in other instances: similar instances yield similar samples. A particular form of sample coordination, the PRN (Permanent Random Numbers) method, was used in survey sampling for almost four decades for Poisson [3] and order [37, 34, 36] samples.

Coordination was (re-)introduced in computer science [5, 4, 9, 25, 26, 2, 13, 27, 14, 17] to address challenges of massive data sets and to facilitate tighter estimates of aggregates over multiple instances. Initially, for $0/1$ values (where instances are sets and sum aggregates of multi-instance functions correspond to set operations) and recently [17] for weighted data.

A particular form of coordination, applicable to bottom-$k$ and Poisson samples is through *consistent ranks*. Rank assignments $\mathbf{r}_i$ of different instances are consistent if for each key $h$, $v_i(h) \geq v_j(h) \Rightarrow r_i(h) \leq r_j(h)$ (in particular, if entries are equal then so are the ranks.) One can get consistent ranks by sharing the seed $u(h) \in U[0,1]$ for a particular key $h$ across instances [17]. This is easily achieved if seeds are determined by random hash functions. For each instance $i$ and key $h$, we assign the rank value $r_i(h) \leftarrow \mathbf{F}^{-1}_{v_i(h)}(u(h))$. For PPS ranks, $r_i(h) = u(h)/v_i(h)$ and for EXP ranks, is $r_i(h) = -\ln(1-u(h))/v_i(h)$.

On decomposable queries, however, coordination results in larger variance than independent samples, due to strong positive correlation between samples of different instances. Thus, independent sampling is preferable when the query workload is dominated by decomposable queries. Coordination also results in unbalanced burden – the same keys are consistently sampled. This is a negative when sampling is used to limit transmissions to save sensor battery power.

**Knowledge of random seeds.** We can get better estimates if we know the random seeds when we compute the estimator. This is possible (without the overhead of incorporating them in the sample) with random rank based weighted sampling, if the seed $u_i(h)$ for instance $i$ are determined by a random hash function of the key $h$. The knowledge of the seed allows the estimator to obtain some information (upper bound) on the value $v_i(h)$ even if it is not sampled. For example in Poisson sampling we know that $v_i(h)$ must satisfy that $\tau < \mathbf{F}^{-1}_{v_i(h)}(u_i(h))$. Since for weighted sampling we have that $F_w$ dominates $F_{w'}$ for $w > w'$ this gives an upper bound on $v_i(h)$. With bottom-$k$ sample, we define $\tau$ to be the $(k+1)$st smallest rank in $K$ and also obtain a similar upper bound when $v_i(h)$ is not sampled.

With coordinated sampling, $u_i(h)$ must be hash generated and reproducible, since they are available for summarization of different instances. With independent sampling of instances, implementations may also use reproducible seeds. We show that knowledge of seeds enhances estimation scope and accuracy of some multi-instance functions.

# 8. APPLICATIONS TO SUM AGGREGATES

## 8.1 Distinct count

Consider two instances with binary values. Each instance can be viewed as a subset of all possible key values $K$, including all keys that have value 1. We are interested in the size of the union of the two sets, that is, the number of distinct keys that occur in at least one instance. The distinct count is a sum aggregate with the function $\text{OR}(v_1(h), v_2(h))$.

Suppose sampling of instances is independent with known seeds: The sampling of each instance can be Poisson or bottom-$k$ but the random seeds used are independent across instances. We estimate the sum by applying the estimators of Section 5.1 to each key, and summing the estimates. As a side note, recall that more accurate estimates are possible by coordinating the samples of different instances, but coordination may not be possible or desirable in some situations. Also recall that we show in Section 6 that if seeds are not known, there is no nonnegative unbiased estimator.

As a motivating application consider two periodic logs of resource requests. Each time period (instance) $i = 1,2$ has a set $N_i$ of active resources (say, resources requested at least once). The set $N_i$ is then summarized via Poisson or bottom-$k$ sampling using random seeds $u_i(h)$ and sampling probability $p_i$, to obtain a sample $S_i$. For Poisson sampling, for all $h \in N$ we have $h \in S_i \iff u_i(h) < p_i$. For bottom-$k$ sampling, $S_i$ includes the $k$ keys in $N_i$ with smallest $u_i(h)$ values. In this case, we use the $(k+1)$st smallest $u_i(h)$ for $p_i$. The random hash functions are such that $u_i(h)$ are independent for $i = 1,2$.

From the samples $S_1$ and $S_2$, and having access to $u_i$ and $p_i$, we want to estimate $D_A = |(N_1 \cup N_2) \cap A|$, the number of distinct keys in $N_1$ and $N_2$ that satisfy some selection rule $A$.

To apply the estimators in Section 5.1, we first categorize sampled keys according to the information we have on their membership in $N_1$ and $N_2$.

$$\begin{aligned}
h \in F_{1?} &\iff h \in S_1 \wedge u_2(h) > p_2 \\
h \in F_{?1} &\iff h \in S_2 \wedge u_1(h) > p_1 \\
h \in F_{11} &\iff h \in S_1 \cap S_2 \\
h \in F_{10} &\iff h \in S_1 \wedge u_2(h) < p_2 \\
h \in F_{01} &\iff h \in S_2 \wedge u_1(h) < p_1
\end{aligned}$$

The HT estimate and variance are

$$\begin{aligned}
\widehat{D_A}^{(HT)} &= \frac{|A \cap (F_{11} \cup F_{10} \cup F_{01})|}{p_1 p_2} . \\
\text{VAR}[\widehat{D_A}^{(HT)}] &= |D_A| \left(\frac{1}{p_1 p_2} - 1\right)
\end{aligned}$$

The L estimate is

$$\widehat{D_A}^{(L)} = \frac{\left|A \cap (F_{1?} \cup F_{?1} \cup F_{11})\right|}{p_1 + p_2 - p_1 p_2} + \frac{\left|A \cap F_{10}\right|}{p_1(p_1 + p_2 - p_1 p_2)} + \frac{\left|A \cap F_{01}\right|}{p_2(p_1 + p_2 - p_1 p_2)}$$

The variance is

$$\text{VAR}[\widehat{D_A}^{(L)}] = |D_A| J_A \text{VAR}[\hat{\text{OR}}^{(L)}|(1,1)] + |D_A|(1-J_A)\text{VAR}[\hat{\text{OR}}^{(L)}|(1,0)] ,$$

where $J_A = \frac{|N_1 \cap N_2 \cap A|}{|(N_1 \cup N_2) \cap A|}$ is the Jaccard coefficient.

We assume in the rest of this section that $p_1 = p_2 = p$ and that we want to estimate the size of the entire union, that is $A = N_1 \cup N_2$. We also assume that $|N_1| = |N_2| = n$. Figure 6 shows the sample size $s$ (which is proportional to the sampling probability $p$) as a function of $n$, for the HT and L estimators. We show this dependency for fixed values of the Jaccard coefficient (denoted $J$ in the figure), and the coefficient of variation (cv – ratio of standard deviation and the mean).

The L estimators allows us to use a smaller sample size (factor of two). When we know that $J$ is above a certain value, we can obtain tighter confidence intervals.

Asymptotically, for small sampling probability $p$, if $N = |N_1 \cup N_2|$ the variance of the HT estimator is $N/p^2$ and its coefficient of variations is $1/(p\sqrt{N})$, meaning that we need to have $p \gg 1/\sqrt{N}$ for meaningful estimates. The variance of the L estimator is $\frac{(1-J)N}{4p^2} + \frac{JN}{2p}$. If $p < \frac{1-J}{2J}$, the coefficient of variation is about $\sqrt{1-J}/(2p\sqrt{N})$, meaning that we need a factor of $\sqrt{1-J}/2$ fewer samples than the HT estimator for the same accuracy. If $p > \frac{1-J}{2J}$, the coefficient of variation is about $\sqrt{J/(2pN)}$, meaning that $\Theta(1)$ samples suffice for any fixed coefficient of variations.

## 8.2 Max dominance

We demonstrate performance of our estimators on a data sets which includes hourly summaries of IP traffic, in the form of destination IP address and number of active IP flows to that destination.

Figure 7 shows performance of $\hat{\max}^{(L)}$ and $\hat{\max}^{(HT)}$ estimators. Instances were from two consecutive hours, each with about $2.45 \times 10^4$ distinct destination IP addresses, with a total of $3.8 \times 10^4$ distinct destinations. The total number of flows in each hour was $5.5 \times 10^5$ and the sum of the maximum values was $7.47 \times 10^5$. The figure shows the normalized variance

$$\frac{\text{VAR}[\sum \hat{\max}]}{(\sum \max)^2} \equiv \frac{\sum \text{VAR}[\hat{\max}]}{(\sum \max)^2}$$

as a function of percentage of sampled keys. The sampling method applied to each instance was PPS Poisson (results are same for priority sampling) and instances were sampled independently but with known seeds. The estimator $\hat{\max}^{(HT)}$ is monotone but not dominant. The estimator $\hat{\max}^{(L)}$ is monotone and dominant. The ratio between the variances of the two estimators on this data set $\text{VAR}[\sum \hat{\max}^{(HT)}]/\text{VAR}[\sum \hat{\max}^{(L)}]$ varied between 2.45 to 2.7.

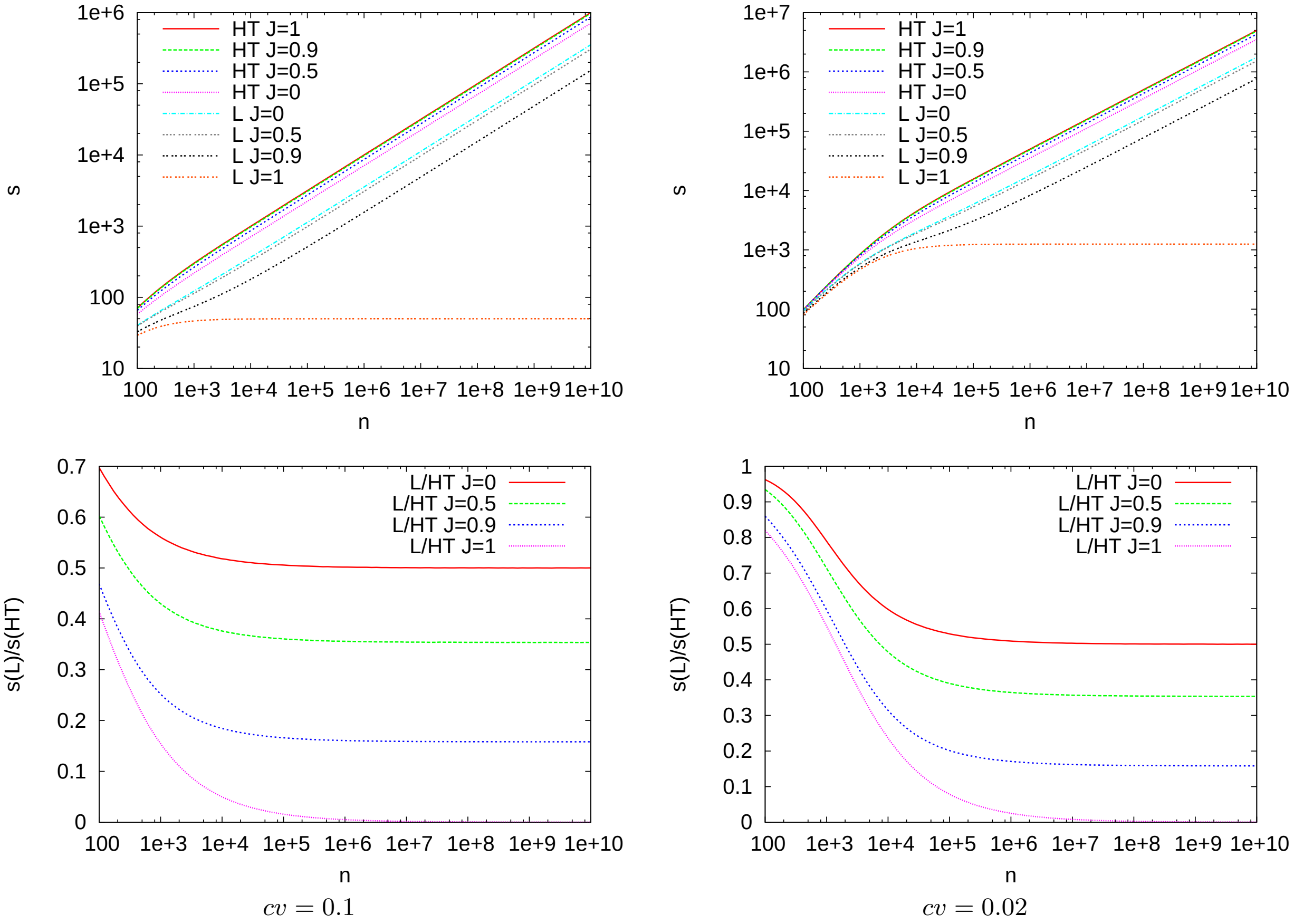

**Figure 6: Sample size $s$ as a function of the input size $n$ (top) and ratio of sample sizes when using the L and HT estimators (bottom) required to achieve certain accuracy (measured by cv).**

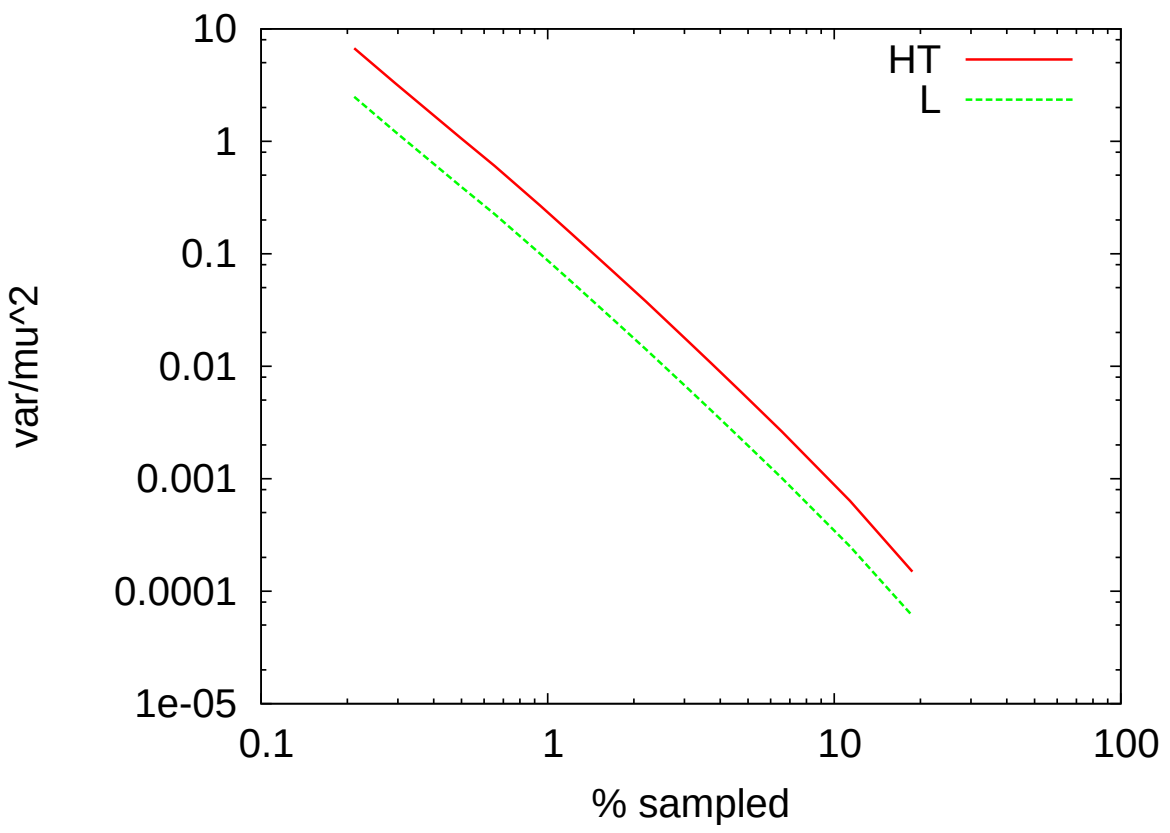

**Figure 7: Variance (normalized) for estimating** $\max$ **dominance using the** $\text{HT}$ **and L estimators over two independently-sampled instances with known seeds (Poisson PPS or priority sampling).**

## Related work

A related and well studied model, not mentioned in the body of the paper, is where data appears as a stream of keys and values, over which we want to estimate frequency moments and $L_p$ norms [24, 1, 30], aiming for query-specific space and time efficient algorithm. Our setup is fundamentally different as the input is a sample base summary of the data and the aim is to design good estimators for different queries.

## Conclusion

Our work laid the foundations for deriving optimal estimators for queries spanning multiple sampled instances. We demonstrated significant improvements over existing estimators for example queries over common sampling schemes. In follow up work, we derive estimators when samples of different instances are coordinated and derive $L_p$ distance estimators over independent and coordinated samples. In the longer run, we hope that sometimes tedious derivations of estimators can be replaced by automated tools.

# APPENDIX

## A. $\hat{\max}^{(L)}$ FOR INDEPENDENT WEIGHTED SAMPLES WITH KNOWN SEEDS

The minimum element of $\prec$ is $\mathbf{0}$, and hence $\mathbf{0}$ is the determining vector of all outcomes consistent with $\mathbf{0}$, which are all outcomes with $S=\emptyset$. Hence, on empty outcomes, $\hat{\max}^{(L)}(S)=0$. We next process vectors with two equal entries $(v,v)$. The outcomes determined by $(v,v)$ are: $S=\{1,2\}$ and $v_1=v_2=v$, $S=\{1\}$ $v_1=v$, and $u_2\geq v_1/\tau_2^*$, or $S=\{2\}$, $v_2=v$, and $u_1\geq v_2/\tau_1^*$. That is, outcomes where both entries are sampled and have the same value $v$ or when exactly one entry is sampled, its value is $v$, and the upper bound on the value of the other entry is at least $v$. The probability of an outcome determined by $(v,v)$ for data $(v,v)$ is $\min\{1,\frac{v}{\tau_1^*}\}+(1-\min\{1,\frac{v}{\tau_1^*}\})\min\{1,\frac{v}{\tau_2^*}\}$. The estimate is therefore

$$\hat{\max}^{(L)}(v,v)=\frac{v}{\min\{1,\frac{v}{\tau_1^*}\}+(1-\min\{1,\frac{v}{\tau_1^*}\})\min\{1,\frac{v}{\tau_2^*}\}}\ . \tag{25}$$

It remains to define the estimator on outcomes that are consistent with data vectors with two different valued entries and not consistent with data vectors with two identical entries: When $|S|=2$ and $v_1\neq v_2$, when $S=\{1\}$ and $u_2\tau_2^*<v_1$ or when $S=\{2\}$ and $u_1\tau_1^*<v_2$. We formulate a system of equations relating the estimate value for determining vectors of the form $(v,v-\Delta)$ $(\Delta\geq 0)$ to the estimate value on determining vectors of the same form and smaller values of $\Delta$. The case of determining vectors of the form $(v-\Delta,v)$ is symmetric.

**case:** $v-\Delta\geq\tau_2^*$: Outcomes consistent with $(v,v-\Delta)$ are $S=\{1,2\}$, in which case the determining vector is $(v,v-\Delta)$, or $S=\{2\}$ and $u_1\tau_1^*>v\geq v-\Delta$, in which case the determining vector is $(v-\Delta,v-\Delta)$. The probability of $S=\{1,2\}$ when the data is $(v,v-\Delta)$ is $\min\{1,\frac{v}{\tau_1^*}\}$. The probability of $S=\{2\}$ is $1-\min\{1,\frac{v}{\tau_1^*}\}$. From Line 13

$$\begin{aligned} v &= \hat{\max}^{(L)}(v,v-\Delta)\min\{1,\frac{v}{\tau_1^*}\}+\\ &\quad \hat{\max}^{(L)}(v-\Delta,v-\Delta)(1-\min\{1,\frac{v}{\tau_1^*}\})\ . \end{aligned}$$

Using (25), $\hat{\max}^{(L)}(v-\Delta,v-\Delta)=v-\Delta$: Substituting and solving for $\hat{\max}^{(L)}(v,v-\Delta)$ we obtain

$$\hat{\max}^{(L)}(v,v-\Delta)=v-\Delta+\frac{\Delta}{\min\{1,\frac{v}{\tau_1^*}\}}\ . \tag{26}$$

**case:** $v\geq\tau_1^*$: Outcomes consistent with data $(v,v-\Delta)$ have $S=\{1,2\}$ or $S=\{1\}$. Outcomes with $S=\{1,2\}$ have determining vector $(v,v-\Delta)$ and probability $\min\{1,\frac{v-\Delta}{\tau^*}\}$. Outcomes with $S=\{1\}$ and $u_2\tau_2^*\geq v$ have determining vector $(v,v)$, estimate value $v$ (using (25)), and probability $(1-\min\{1,\frac{v}{\tau_2^*}\})$. Outcomes with $S=\{1\}$ and $v-\Delta\leq u_2\tau_2^*\leq v$ have determining vector $(v,u_2\tau_2^*)$ , and probability $\frac{\min\{\tau_2^*,v\}-v+\Delta}{\tau_2^*}$.

The equation in Line 13 is

$$\begin{aligned} v &= \hat{\max}^{(L)}(v,v)(1-\min\{1,\frac{v}{\tau_2^*}\})+\\ &\quad \frac{1}{\tau_2^*}\int_{v-\Delta}^{\min\{v,\tau_2^*\}}\hat{\max}^{(L)}(v,y)dy+\\ &\quad \min\{1,\frac{v-\Delta}{\tau_2^*}\}\hat{\max}^{(L)}(v,v-\Delta)\ . \end{aligned}$$

Substituting $\hat{\max}^{(L)}(v,v)=v$, we obtain that $\hat{\max}^{(L)}(v,y)=v$ for all $0\le y\le v$ is a solution.

**case:** $\tau_2^* > v-\Delta,\ \tau_1^* > v$

$$\begin{aligned} v &= \hat{\max}^{(L)}(v,v-\Delta)\frac{v}{\tau_1^*}\frac{v-\Delta}{\tau_2^*} + \\ &\quad \hat{\max}^{(L)}(v,v)\frac{v}{\tau_1^*}(1-\min\{1,\frac{v}{\tau_2^*}\}) + \\ &\quad \hat{\max}^{(L)}(v-\Delta,v-\Delta)(1-\frac{v}{\tau_1^*})\frac{v-\Delta}{\tau_2^*} + \\ &\quad \frac{v}{\tau_1^*}\frac{1}{\tau_2^*}\int_{v-\Delta}^{\min\{v,\tau_2^*\}} \hat{\max}^{(L)}(v,y)dy + \end{aligned} \tag{27}$$

The first term is for outcomes with $S=\{1,2\}$. The determining vector is $(v,v-\Delta)$ and the probability given data vector $(v,v-\Delta)$ is $\frac{v}{\tau_1^*}\frac{v-\Delta}{\tau_2^*}$. The second is when $S=\{1\}$ and $u_2\tau_2^*\ge v$, that is, the upper bound on $v_2$ is at least $v$. The determining vector of these outcomes is $(v,v)$. The third is when $S=\{2\}$ and $u_1\tau_1^*\ge v$, that is, the upper bound on the first entry is at least $v$. The determining vector of these outcomes is $(v-\Delta,v-\Delta)$. The fourth is when $S=\{1\}$ and the upper bound on the second entry is $y\in[v-\Delta,\min\{v,\tau_2^*\}]$. The determining vector is $v,y$. The second term is zero if $\tau_2^*<v$.

We solve separately for two subcases.

**subcase** $\tau_1^*,\tau_2^*\ge v$: We simplify (27)

$$\begin{aligned} v &= \hat{\max}^{(L)}(v,v-\Delta)\frac{v}{\tau_1^*}\frac{v-\Delta}{\tau_2^*} + \\ &\quad \hat{\max}^{(L)}(v,v)\frac{v}{\tau_1^*}(1-\frac{v}{\tau_2^*}) + \\ &\quad \hat{\max}^{(L)}(v-\Delta,v-\Delta)(1-\frac{v}{\tau_1^*})\frac{v-\Delta}{\tau_2^*} + \\ &\quad \frac{v}{\tau_1^*\tau_2^*}\int_{v-\Delta}^{v} \hat{\max}^{(L)}(v,y)dy \end{aligned}$$

We apply (25) to obtain:

$$\begin{aligned} \hat{\max}^{(L)}(v,v) &= \frac{\tau_1^*\tau_2^*}{\tau_1^*+\tau_2^*-v} \\ \hat{\max}^{(L)}(v-\Delta,v-\Delta) &= \frac{\tau_1^*\tau_2^*}{\tau_1^*+\tau_2^*-v+\Delta}\ . \end{aligned}$$

Substituting, we obtain

$$\begin{aligned} v &= \hat{\max}^{(L)}(v,v-\Delta)\frac{v(v-\Delta)}{\tau_1^*\tau_2^*} + \\ &\quad \frac{\tau_1^*\tau_2^*}{\tau_1^*+\tau_2^*-v}\frac{v}{\tau_1^*}(\frac{\tau_2^*-v}{\tau_2^*}) + \\ &\quad \frac{\tau_1^*\tau_2^*}{\tau_1^*+\tau_2^*-v+\Delta}\frac{\tau_1^*-v}{\tau_1^*}\frac{v-\Delta}{\tau_2^*} + \\ &\quad \frac{v}{\tau_1^*\tau_2^*}\int_{v-\Delta}^{v} \hat{\max}^{(L)}(v,y)dy \end{aligned}$$

$$\begin{aligned} v &= \hat{\max}^{(L)}(v,v-\Delta)\frac{v(v-\Delta)}{\tau_1^*\tau_2^*} + \\ &\quad \frac{v(\tau_2^*-v)}{\tau_1^*+\tau_2^*-v} + \frac{(\tau_1^*-v)(v-\Delta)}{\tau_1^*+\tau_2^*-v+\Delta} + \\ &\quad \frac{v}{\tau_1^*\tau_2^*}\int_{v-\Delta}^{v} \hat{\max}^{(L)}(v,y)dy \end{aligned}$$

We define for $\Delta\ge x\ge 0$, $g(x)\equiv \hat{\max}^{(L)}(v,v-x)$ and $G(x)=\int g(x)dx$. Rewriting the above, we obtain

$$\begin{aligned} v &= g(\Delta)\frac{v(v-\Delta)}{\tau_1^*\tau_2^*} + \frac{v(\tau_2^*-v)}{\tau_1^*+\tau_2^*-v} + \\ &\quad \frac{(\tau_1^*-v)(v-\Delta)}{\tau_1^*+\tau_2^*-v+\Delta} + \frac{v}{\tau_1^*\tau_2^*}(G(\Delta)-G(0)) \end{aligned}$$

Taking a partial derivative with respect to $\Delta$

$$\begin{aligned} 0 &= \frac{\partial g(\Delta)}{\partial\Delta}\frac{v(v-\Delta)}{\tau_1^*\tau_2^*} - \frac{v}{\tau_1^*\tau_2^*}g(\Delta) - \\ &\quad \frac{(\tau_1^*-v)(\tau_1^*+\tau_2^*)}{(\tau_1^*+\tau_2^*-v+\Delta)^2} + \frac{v}{\tau_1^*\tau_2^*}g(\Delta) \end{aligned}$$

$$\frac{\partial g(\Delta)}{\partial\Delta} = \frac{\tau_1^*\tau_2^*(\tau_1^*-v)(\tau_1^*+\tau_2^*)}{v(\tau_1^*+\tau_2^*-v+\Delta)^2(v-\Delta)}$$

We use $g(0)=\frac{\tau_1^*\tau_2^*}{\tau_1^*+\tau_2^*-v}$ and the derivative to determine $g(\Delta)\equiv \hat{\max}^{(L)}(v,v-\Delta)$:

$$\begin{aligned} &\hat{\max}^{(L)}(v,v-\Delta) = \\ &= g(\Delta) = g(0) + \int_0^{\Delta}\frac{\partial g(x)}{\partial x}dx \\ &= \frac{\tau_1^*\tau_2^*}{\tau_1^*+\tau_2^*-v} + \frac{\tau_1^*\tau_2^*(\tau_1^*-v)(\tau_1^*+\tau_2^*)}{v}\ \cdot \\ &\quad \cdot\int_0^{\Delta}\frac{1}{(\tau_1^*+\tau_2^*-v+x)^2(v-x)}dx \end{aligned}$$

Integrating[2] we obtain

$$\begin{aligned} &\int_0^{\Delta}\frac{1}{(\tau_1^*+\tau_2^*-v+x)^2(v-x)}dx = \\ &= \frac{1}{(\tau_1^*+\tau_2^*)^2}\ln\left(\frac{(\tau_1^*+\tau_2^*-v+\Delta)v}{(v-\Delta)(\tau_1^*+\tau_2^*-v)}\right) + \\ &+ \frac{\Delta}{(\tau_1^*+\tau_2^*)(\tau_1^*+\tau_2^*-v+\Delta)(\tau_1^*+\tau_2^*-v)} \end{aligned} \tag{28}$$

Substituting:

$$\begin{aligned} &\hat{\max}^{(L)}(v,v-\Delta) = \\ &= \frac{\tau_1^*\tau_2^*}{\tau_1^*+\tau_2^*-v} + \\ &+ \frac{\tau_1^*\tau_2^*(\tau_1^*-v)}{v(\tau_1^*+\tau_2^*)}\ln\left(\frac{(\tau_1^*+\tau_2^*-v+\Delta)v}{(v-\Delta)(\tau_1^*+\tau_2^*-v)}\right) + \\ &+ \frac{\Delta\tau_1^*\tau_2^*(\tau_1^*-v)}{v(\tau_1^*+\tau_2^*-v+\Delta)(\tau_1^*+\tau_2^*-v)} \end{aligned} \tag{29}$$

**subcase** $\tau_1^* > v > \tau_2^* > v-\Delta$: Simplifying (27)

$$\begin{aligned} v &= \hat{\max}^{(L)}(v,v-\Delta)\frac{v(v-\Delta)}{\tau_1^*\tau_2^*} + \\ &\quad \hat{\max}^{(L)}(v-\Delta,v-\Delta)\frac{(\tau_1^*-v)(v-\Delta)}{\tau_1^*\tau_2^*} + \\ &\quad \frac{v}{\tau_1^*\tau_2^*}\int_{v-\Delta}^{\tau_2^*} \hat{\max}^{(L)}(v,y)dy + \end{aligned}$$

[2] We change variables: $y=\tau_1^*+\tau_2^*-v+x$. Then, $v-x=\tau_1^*+\tau_2^*-y$. Integral becomes $\int_{\tau_1^*+\tau_2^*-v}^{\tau_1^*+\tau_2^*-v+\Delta}\frac{1}{y^2(\tau_1^*+\tau_2^*-y)}dy$. We use $B=\tau_1^*+\tau_2^*$. We have (in the range $y\in(0,B)$): $\int\frac{1}{y^2(B-y)}dy = B^{-2}\ln(\frac{y}{B-y})-(By)^{-1}$.

We substitute, using (25):

$$\hat{\max}^{(L)}(v-\Delta, v-\Delta) = \frac{\tau_1^*\tau_2^*}{\tau_1^*+\tau_2^*-v+\Delta}$$

We obtain

$$\begin{aligned} v &= \hat{\max}^{(L)}(v, v-\Delta)\frac{v(v-\Delta)}{\tau_1^*\tau_2^*} + \\ & \frac{\tau_1^*\tau_2^*}{\tau_1^*+\tau_2^*-v+\Delta}\frac{(\tau_1^*-v)(v-\Delta)}{\tau_1^*\tau_2^*} + \\ & \frac{v}{\tau_1^*\tau_2^*}\int_{v-\Delta}^{\tau_2^*} \hat{\max}^{(L)}(v,y)dy + \end{aligned}$$

Simplifying, and using $g(x) \equiv \hat{\max}^{(L)}(v, v-x)$ and $G(x) = \int g(x)dx$:

$$\begin{aligned} v = & \\ & g(\Delta)\frac{v(v-\Delta)}{\tau_1^*\tau_2^*} + \frac{(\tau_1^*-v)(v-\Delta)}{\tau_1^*+\tau_2^*-v+\Delta} + \\ & \frac{v}{\tau_1^*\tau_2^*}(G(\Delta)-G(v-\tau_2^*)) + \end{aligned}$$

We taking a partial derivative with respect to $\Delta$:

$$\begin{aligned} 0 &= g'(\Delta)\frac{v(v-\Delta)}{\tau_1^*\tau_2^*} - \frac{v}{\tau_1^*\tau_2^*}g(\Delta) + \\ & -\frac{(\tau_1^*-v)(\tau_1^*+\tau_2^*)}{(\tau_1^*+\tau_2^*-v+\Delta)^2} + \frac{v}{\tau_1^*\tau_2^*}g(\Delta) \end{aligned}$$

Simplifying,

$$g'(\Delta) = \frac{(\tau_1^*\tau_2^*)(\tau_1^*-v)(\tau_1^*+\tau_2^*)}{(\tau_1^*+\tau_2^*-v+\Delta)^2 v(v-\Delta)}$$

Thus

$$g(\Delta) = g(v-\tau_2^*) + \int_{v-\tau_2^*}^{\Delta} g'(x)dx \ .$$

Using (26), $g(v-\tau_2^*) = \hat{\max}^{(L)}(v, \tau_2^*) = \tau_1^*+\tau_2^*-\frac{\tau_1^*\tau_2^*}{v}$. Hence,

$$\begin{aligned} & \hat{\max}^{(L)}(v, v-\Delta) = \\ & \quad \tau_1^*+\tau_2^*-\frac{\tau_1^*\tau_2^*}{v} + \\ & \quad \frac{(\tau_1^*\tau_2^*)(\tau_1^*-v)(\tau_1^*+\tau_2^*)}{v}\int_{v-\tau_2^*}^{\Delta}\frac{1}{(\tau_1^*+\tau_2^*-v+x)^2(v-x)}dx \end{aligned}$$

Using (28),

$$\begin{aligned} & \int_{v-\tau_2^*}^{\Delta}\frac{1}{(\tau_1^*+\tau_2^*-v+x)^2(v-x)}dx \\ & = \frac{1}{(\tau_1^*+\tau_2^*)^2}\ln\left(\frac{(\tau_1^*+\tau_2^*-v+\Delta)\tau_1^*}{\tau_2^*(\tau_1^*+\tau_2^*-v)}\right) + \\ & + \frac{\tau_2^*-v+\Delta}{(\tau_1^*+\tau_2^*)(\tau_1^*+\tau_2^*-v+\Delta)\tau_1^*} \end{aligned}$$

$$\begin{aligned} & \hat{\max}^{(L)}(v, v-\Delta) = \\ & = \tau_1^*+\tau_2^*-\frac{\tau_1^*\tau_2^*}{v} + \\ & + \frac{(\tau_1^*\tau_2^*)(\tau_1^*-v)}{v(\tau_1^*+\tau_2^*)}\ln\left(\frac{(\tau_1^*+\tau_2^*-v+\Delta)\tau_1^*}{\tau_2^*(\tau_1^*+\tau_2^*-v)}\right) \\ & + \frac{\tau_2^*(\tau_1^*-v)(\tau_2^*-v+\Delta)}{(\tau_1^*+\tau_2^*-v+\Delta)v} \end{aligned} \tag{30}$$

The expressions stated in the table in Figure 3 are from (26), (29), and (30).